# A Second Order Radiative Transfer Equation and Its Solution by Meshless Method with Application to Strongly Inhomogeneous Media

J.M. Zhao[a], J.Y. Tan[b], L.H. Liu[a,b]*

[a] *School of Energy Science and Engineering, Harbin Institute of Technology, 92 West Dazhi Street, Harbin 150001, People's Republic of China*

[b] *School of Auto Engineering, Harbin Institute of Technology at Weihai, 2 West Wenhua Road, Weihai 264209, People's Republic of China*

**Abstract**

A new second order form of radiative transfer equation (named MSORTE) is proposed, which overcomes the singularity problem of a previously proposed second order radiative transfer equation [J. Comput. Phys. 214 (2006) 12-40 (where it was termed SAAI), Numer. Heat Transfer B 51 (2007) 391-409] in dealing with inhomogeneous media where some locations have very small/zero extinction coefficient. The MSORTE contains a naturally introduced diffusion (or second order) term which provides better numerical property than the classic first order radiative transfer equation (RTE). The stability and convergence characteristics of the MSORTE discretized by central difference scheme is analyzed theoretically, and the better numerical stability of the second order form radiative transfer equations than the RTE when discretized by the central difference type method is proved. A collocation meshless method is developed based on the MSORTE to solve radiative transfer in inhomogeneous media. Several critical test cases are taken to verify the performance of the presented method. The collocation meshless method based on the MSORTE is demonstrated to be capable of stably and accurately solve radiative transfer in strongly inhomogeneous media, media with void region and even with discontinuous extinction coefficient.

---

\* Corresponding author. Tel.: +86-451-86402237; fax: +86-451-86221048.
*Email addresses:* jmzhao@hit.edu.cn, tanjy@hit.edu.cn, lhliu@hit.edu.cn

**Nomenclature**

| | |
|---|---|
| **B** | Tool matrix defined in Eq. (53a) |
| **b** | Tool vector defined in Eq. (53b) |
| $C_l$ | Expansion coefficient |
| **D** | Differential matrix defined in Eq. (53d)-(h) |
| $D$ | Diameter, m |
| $d$ | Length scale of inclusion, m |
| $G$ | Incident radiation, W/m$^2$ |
| **h** | Vector defined in Eq. (51) |
| $I$ | Radiative intensity, W/m$^2$sr |
| **K** | Stiff matrix defined in Eq. (51) |
| $L$ | Length of ray trajectory, side length of square enclosure |
| **n** | Normal vector |
| $N_{sol}$ | Total number of solution nodes |
| $N$ | Nodal basis function |
| $P_l$ | Legendre polynomial of $l$-th order |
| $q$ | Radiative heat flux, W/m$^2$ |
| $s$ | Ray trajectory coordinates |
| **s** | Vector defined in Eq. (53c) |
| $S$ | Source function defined in Eq. (3) |
| $T$ | Temperature, K |
| $T^*$ | False Temperature $T^* = \sqrt[4]{G/4\sigma}$, K |
| $U$ | Unit step function |
| $V$ | Solution domain |
| $w$ | MLS weight function |
| $\bar{w}$ | Angular quadrature weight |
| $\tilde{w}$ | Radial basis function defined by Eq. (48) |
| $\mathbf{x}, \bar{\mathbf{x}}$ | Spatial coordinates vector |
| $x, y, z$ | Cartesian coordinates |
| $\alpha$ | Amplifying factor |
| $\beta$ | Extinction coefficient, m$^{-1}$ |
| $\delta$ | Kronecker delta |
| $\varepsilon_w$ | Wall emissivity |
| $\varphi$ | Azimuth angle |
| $\Gamma$ | Domain boundary |
| $\Phi$ | Scattering phase function |
| $\tilde{\Phi}$ | Modified scattering phase function |
| $\kappa_a$ | Absorption coefficient, m$^{-1}$ |
| $\kappa_s$ | Scattering coefficient, m$^{-1}$ |
| $\mu, \xi, \eta$ | Cartesian components of $\mathbf{\Omega}$ |
| $\varpi, \bar{\varpi}$ | Frequency domain variable, reduced frequency |
| $\theta$ | Zenith angle |
| $\sigma$ | Stefan-Boltzmann constant, W/m$^2$K$^4$ |
| $\tau$ | Optical thickness |
| $\tau_\Delta$ | Grid optical thickness |
| $\omega$ | Single scattering albedo |
| $\mathbf{\Omega}, \mathbf{\Omega}'$ | Vector of radiation direction |
| $\Omega, \Omega'$ | Solid angle |

*Subscripts and Superscripts*

| | |
|---|---|
| 0 | Inflow |
| b | Black body |
| g | Medium |
| I | Inclusion |
| i, j | Spatial solution node index |
| m | The *m*th angular direction |
| w | Value at wall |
| $\varphi$ | Azimuth |
| $\theta$ | Zenith |

## 1. Introduction

Radiative heat transfer in absorbing, emitting, and scattering media is important in many scientific and engineering disciplines. The classic governing equation of steady radiative transfer (RTE) can be written simply as [1]

$$\mathbf{\Omega} \cdot \nabla I + \beta I = S \tag{1}$$

where $\mathbf{\Omega} = \mu\mathbf{i} + \eta\mathbf{j} + \xi\mathbf{k}$ is the unit direction vector of radiation, $\beta$ is the extinction coefficient, $S$ is the source term accounting for the thermal emission of medium and in scattering. Here the angular integral part of the RTE is suppressed into the source term. As is seen, the RTE [Eq.(1)] is a first order differential equation. The first term of the left hand side of Eq. (1) can be seen as a convection term with a convection velocity of $\mathbf{\Omega}$, namely, $\mu$, $\eta$ and $\xi$ are taken as the velocity in x-, y- and z- directions, respectively. As compared to the convection diffusion equation, the RTE can be considered as a special kind of convection-dominated equation [2] without the diffusion term. The convection-dominated property of an equation may cause nonphysical oscillation in the numerical results. This type of instability occurs in many numerical methods, such as finite difference methods, finite element methods and meshless methods, etc., if no special stability treatment is taken. Furthermore, the instability caused by convection-dominated characteristics is often coupled with other numerical errors, such as 'ray effects' [3-8].

Many numerical methods have been developed to solve radiative transfer in semitransparent media in recent years. These methods can be mainly classified into two groups according to their basic principle, (1) methods based on ray tracing, such as ray tracing method [9, 10], zonal method [11], Monte Carlo method [12-14], and discrete transfer method, and (2) methods based discretization of partial differential equations, such as discrete ordinate methods (DOM) [15-17], finite volume method (FVM) [18-20] and finite element method (FEM) [21-23]. The numerical properties of the RTE is affected little by the ray tracing based methods because the simulation process of these methods is more physically based and the ray path is analytically determined, which does not explicitly rely on the differential form of the RTE. However, these methods have some important drawbacks, such as, often being time consuming even for relatively simple problems, being difficult to deal with anisotropic scattering, and being difficult to be implemented to complex domains due to the geometrical shielding. As a complement, the partial differential equation discretization based methods have advantages of ease, efficiency and flexibility to deal with multidimensional complex radiative transfer problems, and have received considerable attention in recent years. However, these methods suffer much from the instability caused by the convection-dominated property of the RTE.

As for the second group of methods, in order to make the numerical discretization schemes correctly model the transfer process, special stabilization techniques, such as upwinding scheme or artificial diffusion, are used often in FVM, DOM and FEM. Besides taking various numerical stabilization schemes, another way to circumvent this stability problem is to analytically transform the RTE into a numerically stable equation, for example, the second order partial differential equation. The second order derivative term is known to have the characteristic of diffusion and good numerical properties. The diffusion term introduced in the analytical transformation process is natural and consistent with original first order equation, which is distinctly different than the stabilization techniques mentioned previously. As compared to the different special stabilization techniques, which is often not easy to be designed for different methods, the transformed second order equation can be solved stably with standard numerical methods. Hence this can be considered a unified approach to overcome the stability problem for different numerical methods. As will be demonstrated in this paper, the meshless method is applied to a new proposed second order RTE to obtain stable solutions.

Currently, two different transformed equations have been proposed. One is the even-parity (EPRTE) formulation of the RTE, which is a second order differential equation of the even parity of radiative intensity. The EPRTE was initially proposed in the field of neutran transport and has been used for decades [24-26]. In recent years, the equation was also applied to radiative heat transfer calculations. Cheong and Song [27] and Liu and Chen [28] studied the DOM solution of the EPRTE. Fiveland and Jessee [29, 30] studied the FEM solution of the EPRTE. Sadi et al [31, 32] studied the meshless method solution of the EPRTE. These works demonstrated the numerical stability of the second order equation. Another one is the second order radiative transfer equation (SORTE) [33] proposed recently, which is a second order differential equation of radiative intensity itself. Though similar stability is obtained from the second order term, as compared to the EPRTE, the SORTE uses radiative intensity as solution variable and is more convenient and easier to be applied to complex radiative transfer problems, such as anisotropic scattering, and implemented in numerical methods. It is noted that the SORTE was also proposed separately by Morel et al. [34], which was termed the self-adjoint angular intensity (SAAI) radiative transfer equation in their paper. The SORTE has been applied and evaluated for solving radiative heat transfer by many numerical methods. Zhao and Liu [33, 35] applied finite/spectral element method to solve the SORTE. Hassanzadeh and Raithby [36], and Tan et al. [37] evaluated the performance of the SORTE by using FVM and meshless method, respectively. However, because the existence of the reciprocal of extinction coefficient in the equation, a singularity problem exists for the proposed second order equation in dealing with inhomogeneous media where some locations have very small or null extinction coefficient both for the EPRTE and the SORTE. This limits the further application of these second order equations in solving relevant engineering radiative transfer problems.

Recently, a new class of methods, meshless methods [38], has been applied to solve radiative transfer in participating media [31, 39-42]. The salient feature of meshless methods is that they provide the ways of building approximation space just based on freely scattered nodes without relying on any mesh structure (e.g. elements, faces and sides) that is essential in traditional methods, such as FVM and FEM. The meshless approach is thus very appealing for inverse geometrical design problems and problems with deformable domain. Recent studies [31, 32] revealed that the meshless method based on the first order RTE lacks of stability and the obtained results suffer from spurious oscillations in some condition, while the meshless method based on the EPRTE is stable. The instability of the meshless method based on the RTE is similar to central difference scheme in the solution of convection dominated problem. However, it is not easy to devise the upwinding scheme for the meshless method as for the FVM. The lack of mesh data structure makes it being difficult to obtain the nodes in the upwind direction. Hence the second order form of the RTE can be considered a better basis for the meshless methods other than the classic RTE.

In this paper, a new second order form of radiative transfer equation (named MSORTE) is proposed to overcome the singularity problem of the SORTE in dealing with inhomogeneous media where some locations have very small extinction coefficient. A meshless method based on the moving least square approximation is developed to solve and verify the versatility and performance of the MSORTE in solving radiative transfer in very strongly inhomogeneous media. The paper is organized as follows. Firstly, the MSORTE is introduced. Then the stability and convergence characteristics of the MSORTE discretized by central difference scheme is analyzed. In Section 3, a meshless method discretization of the proposed equation is presented. Finally, numerical verification of the performance of the developed meshless method based on the MSORTE is presented and discussed.

## 2. The Second Order Radiative Transfer Equation

### 2.1. Derivations

The basic idea on derivation of the new second order equations are similar to the derivation of the SORTE [33]. The general equation of transfer for an absorbing, emitting, and scattering medium in ray coordinate is

$$\frac{dI}{ds} + \beta I = S, \qquad s \in [0, L] \tag{2}$$

where $S$ is the *source function* accounting for emission and in-scattering given as

$$S = \kappa_a I_b + \frac{\kappa_s}{4\pi} \int_{4\pi} I(s, \mathbf{\Omega}') \Phi(\mathbf{\Omega}', \mathbf{\Omega}) d\Omega' \tag{3}$$

$L$ is the length of the ray trajectories considered, and subject to the following inflow boundary condition

$$I = I_0, \quad s = 0 \tag{4}$$

where $I_0$ is the inflow radiative intensity. By directly applying the streaming operator $d/ds$ to Eq. (2), which yields

$$\frac{d^2 I}{ds^2} + \frac{d\beta I}{ds} = \frac{dS}{ds} \tag{5}$$

This equation is named here the MSORTE for it is in a form of mixtures of the first and the second order derivatives. By a comparison with the SORTE [33] [Eq. (6)] given below,

$$-\frac{\mathrm{d}}{\mathrm{d}s}\left[\beta^{-1}\frac{\mathrm{d}I}{\mathrm{d}s}\right]+\beta I = S - \frac{\mathrm{d}\beta^{-1}S}{\mathrm{d}s} \tag{6}$$

It is seen that the MSORTE and the SORTE share the common property of a naturally introduced diffusion (second order) term, which ensures good numerical properties for central difference like discretization schemes as will be analyzed in Section 3. However, the appearance of the reciprocal of extinction coefficient in the SORTE causes difficulty in numerical solution for problem with inhomogeneous media where some locations have very small/zero values of extinction coefficient. On the contrary, the MSORTE does not suffer this problem.

As for the second order radiative transfer equations, besides the original inflow boundary condition for the RTE, another boundary condition for the outflow boundary is also needed. Here, the same outflow boundary condition for the SORTE [33] is prescribed for the MSORTE. The boundary conditions for the inflow and the outflow boundaries are given respectively as

$$I = I_0, \quad s = 0 \tag{7}$$

and

$$\frac{\mathrm{d}I}{\mathrm{d}s} + \beta I = S, \quad s = L \tag{8}$$

*2.2. Cartesian coordinates formulation*

In Cartesian coordinates, we have $I(s,\mathbf{\Omega}) = I[s(x,y,z),\mathbf{\Omega}]$, hence

$$\frac{\mathrm{d}I}{\mathrm{d}s} = \frac{\mathrm{d}x}{\mathrm{d}s}\frac{\partial I}{\partial x} + \frac{\mathrm{d}y}{\mathrm{d}s}\frac{\partial I}{\partial y} + \frac{\mathrm{d}z}{\mathrm{d}s}\frac{\partial I}{\partial z} \tag{9}$$

To consider $\mathrm{d}s$ as curve length differential, the coordinates transformation coefficients $\mathrm{d}x/\mathrm{d}s$, $\mathrm{d}y/\mathrm{d}s$ and $\mathrm{d}z/\mathrm{d}s$ are the direction cosines of the transport direction $\mathbf{\Omega}$. As such, Eq. (9) can be written as

$$\frac{\mathrm{d}I}{\mathrm{d}s} = \mathbf{\Omega}\cdot\nabla I \tag{10}$$

Hence the MSORTE [Eq. (5)] can be expressed as

$$(\mathbf{\Omega}\cdot\nabla)^2 I + \mathbf{\Omega}\cdot\nabla[\beta I] = \mathbf{\Omega}\cdot\nabla S \tag{11}$$

The boundary conditions [Eq. (7) and Eq. (8)] can be formulated as

$$I(\mathbf{x}_w, \mathbf{\Omega}) = I_0(\mathbf{x}_w, \mathbf{\Omega}), \quad \mathbf{n}_w \cdot \mathbf{\Omega} < 0 \tag{12a}$$

$$\mathbf{\Omega}\cdot\nabla I + \beta I = S, \quad \mathbf{n}_w \cdot \mathbf{\Omega} \geq 0 \tag{12b}$$

in which $I_0$ is the known intensity distribution at the boundary and $\mathbf{n}_w$ is the outward normal vector at the boundary wall. The boundary condition Eq. (12a) is conventionally called Dirichlet boundary condition or essential boundary condition, and Eq. (12b) the Neumann boundary condition or natural boundary condition. A schematic of the boundary conditions is shown in Figure 1, where the whole boundary $\Gamma = \Gamma_D \cup \Gamma_N$, $\Gamma_D$ and $\Gamma_N$ denotes the inflow boundary and the outflow boundary, respectively.

For the opaque, diffuse emitting and reflecting wall, the inflow radiative intensity is given as

$$I_0(\mathbf{x}_w, \mathbf{\Omega}) = \varepsilon_w I_b(\mathbf{x}_w) + \frac{1-\varepsilon_w}{\pi} \int_{\mathbf{n}_w \cdot \mathbf{\Omega}' > 0} I(\mathbf{x}_w, \mathbf{\Omega}') |\mathbf{n}_w \cdot \mathbf{\Omega}'| d\Omega' \tag{13}$$

where $\varepsilon_w$ is the wall emissivity.

## 3. Stability and Convergence Characteristics Analysis

In this section, the numerical characteristics of different radiative transfer equations discretizing by central difference schemes are studied in a unified analytical framework in frequency domain. As is known, the central difference discretization is equivalent to the Galerkin FEM in discretization of the convection terms [43-45]. The central difference scheme suffers stability problem for convection-dominated convection diffusion equations, which is also true for discretization of the RTE. On 1-D uniform grid, the shape functions used in meshless method, such as those produced by moving least squares approximation [38], are symmetrical around the definition node, which is similar to the shape function used in FEM. This is similar to the central difference scheme, namely, the discretization nodes is selected symmetrically around the solution node. This indicates the meshless method with symmetrical shape functions will have similar performance of stability with the FEM and the central difference scheme in the discretization of the convection term. As such, the stability of meshless method based on different transfer equations is studied by an analysis based on central difference discretizing of the transfer equations for simplicity.

*3.1. Stability analysis*

In this analysis, the medium properties are assumed to be constant and the medium is considered to be non-scattering, which can ease the analysis and catch major characteristics of the equations in general case. First, we consider the RTE. The RTE in ray coordinate can be discretized (at node $n$) through the central difference scheme as

$$\frac{I_{n+1} - I_{n-1}}{2\Delta s} + \beta I_n = S_n \tag{14}$$

where $\Delta s$ is the discretization step, and the superscript denote the node index. To facilitate the analysis, we want to obtain a differential equation which is equivalent to the difference equation, which can be obtained by Taylor expansion of the intensity $I_{n+1}$ and $I_{n-1}$ at node $n$. By this approach, the difference term in Eq. (14) can be written in its equivalent differential form as

$$\frac{I_{n+1} - I_{n-1}}{2\Delta s} = \frac{\mathrm{d}I_n}{\mathrm{d}s} + \frac{1}{2\Delta s}\sum_{k=2}^{\infty}\left[1-(-1)^k\right](\Delta s)^k \frac{1}{k!}\frac{\mathrm{d}^k I_n}{\mathrm{d}s^k} \tag{15}$$

As a result, the equivalent differential equation of Eq. (14) (at node $n$) can be obtained as

$$\frac{\mathrm{d}I}{\mathrm{d}s} + \beta I = S + \frac{1}{2\Delta s}\sum_{k=2}^{\infty}\left[(-1)^k - 1\right](\Delta s)^k \frac{1}{k!}\frac{\mathrm{d}^k I}{\mathrm{d}s^k} \tag{16}$$

where the subscript '$n$' is omitted for brevity. It is seen that some additional terms exist in the discretized equation [Eq. (16)] as compared to the original RTE. To compare the result obtained by Eq. (16) with the exact result obtained by the RTE and conduct an error analysis will give clue on the numerical stability. The error analyses in frequency domain are conducted in the following.

The Fourier transform of the finite difference equation [Eq. (14)] based on its equivalent differential form [Eq. (16)] can be obtained as

$$\left[j\varpi\,\mathrm{sinc}(\Delta s\varpi) + \beta\right]\hat{I} = \hat{S} \tag{17}$$

where $\varpi$ is corresponding frequency domain variable and $j=\sqrt{-1}$. The detailed frequency domain relations during the derivation is presented in Appendix A. It is easily verify that Eq. (17) approaches the RTE when $\Delta s \to 0$. A solution to Eq. (17) can be obtained as

$$\hat{I} = \frac{\hat{S}}{j\varpi \text{sinc}(\Delta s\varpi) + \beta} \tag{18}$$

While an exact solution to the RTE yields

$$\hat{I}_E = \frac{\hat{S}}{j\varpi + \beta} \tag{19}$$

Hence a relative error in frequency domain can be defined and obtained as

$$E_{I,RTE} = \frac{\hat{I} - \hat{I}_E}{\hat{I}_E} = \frac{j\varpi + \beta}{j\varpi \text{sinc}(\Delta s\varpi) + \beta} - 1$$
$$= \frac{j2\pi\bar{\varpi}\tau_\Delta^{-1} + 1}{j2\pi\bar{\varpi}\tau_\Delta^{-1}\text{sinc}(2\pi\bar{\varpi}) + 1} - 1 \tag{20}$$

in which, $\bar{\varpi}$ is a reduced frequency defined as $\bar{\varpi} = \dfrac{\Delta s\varpi}{2\pi}$ and $\tau_\Delta = \beta\Delta s$ is the grid optical thickness.

By using the central difference scheme, the MSORTE can be discretized as follows

$$\frac{I^{n+1} - 2I^n + I^{n-1}}{(\Delta s)^2} + \beta\frac{I^{n+1} - I^{n-1}}{2\Delta s} = \frac{S^{n+1} - S^{n-1}}{2\Delta s} \tag{21}$$

And its Fourier transform can be obtained using the relations presented in Appendix A as

$$\left[-\varpi^2 \text{sinc}^2(\Delta s\varpi/2) + \beta j\varpi \text{sinc}(\Delta s\varpi)\right]\hat{I} = j\varpi \text{sinc}(\Delta s\varpi)\hat{S} \tag{22}$$

Hence

$$\hat{I} = \frac{j\varpi \text{sinc}(\Delta s\varpi)}{-\varpi^2 \text{sinc}^2(\Delta s\varpi/2) + \beta j\varpi \text{sinc}(\Delta s\varpi)}\hat{S} \tag{23}$$

and a relative error in frequency domain for the solution based on the MSORTE can be obtained as

$$E_{I,MSORTE} = \frac{\hat{I} - \hat{I}_E}{\hat{I}_E} = \frac{j\varpi \text{sinc}(\Delta s\varpi)(j\varpi + \beta)}{-\varpi^2 \text{sinc}^2(\Delta s\varpi/2) + \beta j\varpi \text{sinc}(\Delta s\varpi)} - 1$$
$$= \frac{j2\pi\bar{\varpi}\tau_\Delta^{-1} + 1}{j2\pi\bar{\varpi}\tau_\Delta^{-1}\text{sinc}^2(\pi\bar{\varpi})\text{sinc}^{-1}(2\pi\bar{\varpi}) + 1} - 1 \tag{24}$$

Similarly, the SORTE [Eq. (6)] can be discretized as

$$-\frac{I^{n+1} - 2I^n + I^{n-1}}{(\Delta s)^2} + \beta^2 I^n = \beta S^n - \frac{S^{n+1} - S^{n-1}}{2\Delta s} \tag{25}$$

and its Fourier transform can be obtained similarly as

$$\left[\beta^2 + \varpi^2 \text{sinc}^2(\Delta z \varpi / 2)\right]\hat{I} = \beta\hat{S} - j\varpi\,\text{sinc}(\varpi\Delta z)\hat{S} \tag{26}$$

Hence a relative error in frequency domain for the solution based on the SORTE is obtained as

$$\begin{aligned} E_{I,SORTE} &= \frac{\hat{I} - \hat{I}_E}{\hat{I}_E} = \frac{\left[\beta - j\varpi\,\text{sinc}(\varpi\Delta z)\right](\beta + j\varpi)}{\left[\beta^2 + \varpi^2 \text{sinc}^2(\Delta z \varpi / 2)\right]} - 1 \\ &= \frac{\left[1 - j2\pi\bar{\varpi}\tau_\Delta^{-1}\text{sinc}(2\pi\bar{\varpi})\right]\left(1 + j2\pi\bar{\varpi}\tau_\Delta^{-1}\right)}{1 + \left(2\pi\bar{\varpi}\tau_\Delta^{-1}\right)^2 \text{sinc}^2(\pi\bar{\varpi})} - 1 \end{aligned} \tag{27}$$

A comparison of the solution error for the different equations at different grid optical thickness is presented in Figure 2. The frequency range of the reduced frequency $\bar{\varpi}$ is plotted in [0, 0.5]. This is based on the fact that the maximum frequency (or shortest wavelength) of a harmonic that can propagate on a uniform grid of spacing $\Delta s$ is $\pi / \Delta s$ (or wavelength $2\Delta s$), namely, $\bar{\varpi} = 0.5$. For all the transfer equations, it can be seen that the relative error increases with $\bar{\varpi}$ for different grid optical thickness, and the maximum relative error occur at $\bar{\varpi} = 0.5$. The RTE shows a huge relative error at $\bar{\varpi} = 0.5$, with a value greater than 300 for $\tau_\Delta = 0.01$, which is about two order of magnitude greater than that of the MSORTE and the SORTE, which are about 1. Hence for the RTE, significant error can be observed even if the source $\hat{S}$ has a small magnitude near $\bar{\varpi} = 0.5$ because it will be magnified to two order of magnitude. This is true for the case in which the emission field contains large gradient. The large gradient will induce a wide band spectrum distribution according to Fourier analysis. It is also observed that the solution errors (especially at the high frequency) of the obtained by the RTE reduces significantly with the increasing of grid optical thickness $\tau_\Delta$, which indicates the solution error will decrease for problem with larger extinction coefficient on a specified grid. This observation will be demonstrated in Section 5.1.1 for one-dimensional case and Section 5.2.1 for two-dimensional case. With the increasing of grid optical thickness, say for $\tau_\Delta > 1$ (the cell size $\Delta s$ is greater than the mean-free-path $\beta^{-1}$), the performance of the RTE gets comparable with that of the MSORTE and the SORTE. This reveals that the RTE, SORTE and MSORTE have comparable accuracy when the cell size $\Delta s$ is greater than the mean-free-path $\beta^{-1}$.

Detailed relation of the solution error with grid optical thickness for different equations is shown in Figure 3, which plots the relative errors at three different reduced frequencies, $\bar{\varpi} = 0.5$, 0.25 and 0.1. At high frequency, $\bar{\varpi} = 0.5$, the RTE shows huge relative error at small grid optical thickness ($\tau_\Delta < 1$). This indicates that the RTE is numerically unstable for high frequency mode. At medium ($\bar{\varpi} = 0.25$) or low frequency ($\bar{\varpi} = 0.1$), the RTE shows comparable relative error with the MSORTE and the SORTE. As a result, in order to accurately solve a problem in which the intensity contains high frequency modes, the grid size should be reduced too much, such that the $\bar{\varpi}$ is decreased to ensure a reasonable accuracy.

However, for both the low and the high frequency modes, the MSORTE and the SORTE show comparable relative error and are superior to the RTE at small grid optical thickness ($\tau_\Delta < 1$). In practice, the grid optical thickness in the range of $\tau_\Delta < 1$ is the most commonly encountered. Hence the MSORTE and the SORTE is a better choice in solving general radiative transfer problem, especially for the problems in which the intensity contains non-ignorable high frequency modes.

It is noted that there is a cross between the error curve of the RTE and the MSORTE at $\bar{\varpi} = 0.5$ (the top plot in Figure 3), the cross condition can be obtained by setting modulus of Eq. (20) to 1, it gives $\tau_\Delta = \pi$. This indicate the high frequency performance of the RTE will be superior to the MSORTE for grid optical thickness $\tau_\Delta > \pi$.

These theoretical results interpret well the reported numerical stability problem of the RTE solution using Galerkin finite element method [33] in dealing with such kind of problems: (1) the strong spurious wiggles appears in the solution, (2) the spurious wiggles has a short wavelength about two times of the grid spacing and (3) the magnitude of the spurious wiggles tends to decrease with the increasing of extinction coefficient. By comparison with the frequency domain error distribution of the solution based on the RTE, it proves the better numerical stability of the MSORTE and the SORTE.

*3.2. Convergence of source iteration*

Source iteration is a general procedure to solve the discrete-ordinates radiative transfer equation in scattering medium, which is also used in the meshless method solution of this paper. Here the convergence characteristics of the different radiative transfer equations are analyzed based on a similar approach used in Section 3.1.

To consider the scattering effect, angular coordinate need to be included in the formulation. Here the one dimensional RTE is considered, which can be written as

$$\xi \frac{\mathrm{d}I}{\mathrm{d}z} + \beta I = \kappa_a I_b + \frac{\kappa_s}{4\pi} G^* \qquad (28)$$

where the superscript '*' denotes value at previous iteration. In this case, the discretized form in frequency domain, Eq. (18), can be written as

$$\hat{I} = \frac{1}{j\varpi\xi\mathrm{sinc}(\Delta z\varpi) + \beta}\left(\kappa_s \hat{I}_b + \frac{\kappa_s}{4\pi}\hat{G}^*\right) \qquad (29)$$

Integrating Eq. (29) in the angular space, it yields

$$\hat{G} = 2\pi\int_{-1}^{1}\frac{1}{j\xi\varpi\mathrm{sinc}(\varpi\Delta z) + \beta}\mathrm{d}\xi\left[\kappa_s \hat{I}_b + \frac{\kappa_s}{4\pi}\hat{G}^*\right] \qquad (30)$$

The angular integration in Eq. (30) can be explicitly evaluated as

$$\int_{-1}^{1}\frac{1}{j\xi\varpi\mathrm{sinc}(\varpi\Delta z) + \beta}\mathrm{d}\xi = 2\frac{\mathrm{arctg}\left[\varpi\beta^{-1}\mathrm{sinc}(\varpi\Delta z)\right]}{\varpi\mathrm{sinc}(\varpi\Delta z)}$$
$$= 2\beta^{-1}\mathrm{arctgenc}\left[\varpi\beta^{-1}\mathrm{sinc}(\varpi\Delta z)\right] \qquad (31)$$

where $\mathrm{arctgenc}(x) = \mathrm{arctg}(x)/x$ is a function introduced to shorten the formula, which is defined follow the definition of the sine cardinal function (sinc). Substituting Eq. (31) into Eq. (30), it yields

$$\hat{G} = \omega \operatorname{arctgenc}\left[\varpi\beta^{-1}\operatorname{sinc}(\varpi\Delta z)\right]\hat{G}^*$$
$$+ \operatorname{arctgenc}\left[\varpi\beta^{-1}\operatorname{sinc}(\varpi\Delta z)\right]4\pi(1-\omega)\hat{I}_b \tag{32}$$

The relation of errors at each iteration is thus obtained as

$$\hat{E}_G = \omega \operatorname{arctgenc}\left[\varpi\beta^{-1}\operatorname{sinc}(\varpi\Delta z)\right]\hat{E}_G^* \tag{33}$$

where the error is defined as $\hat{E}_G = \hat{G} - \hat{G}_E$, $\hat{G}_E$ is the exact incident radiation in frequency domain. An error amplifying factor can thus be defined (for discretization based on the RTE) as

$$\alpha_{RTE} = \omega \operatorname{arctgenc}\left[\varpi\beta^{-1}\operatorname{sinc}(\varpi\Delta z)\right]$$
$$= \omega \operatorname{arctgenc}\left[2\pi\bar{\varpi}\tau_\Delta^{-1}\operatorname{sinc}(2\pi\bar{\varpi})\right] \tag{34}$$

which is a function of reduced frequency, grid optical thickness $\tau_\Delta = \beta\Delta z$ and scattering albedo $\omega$. The amplifying factor reveals the convergence speed of the scheme. A smaller value indicates fast convergence speed.

Following the similar procedure, the amplifying factor for central difference discretization based on the MSORTE and the SORTE can also be obtained, which are given as follows

$$\alpha_{MSORTE} = \omega \operatorname{arctgenc}\left(\frac{2\pi\bar{\varpi}}{\tau_\Delta}\frac{\operatorname{sinc}^2(\pi\bar{\varpi})}{\operatorname{sinc}(2\pi\bar{\varpi})}\right) \tag{35}$$

$$\alpha_{SORTE} = \omega \operatorname{arctgenc}\left[2\pi\bar{\varpi}\tau_\Delta^{-1}\operatorname{sinc}(\pi\bar{\varpi})\right] \tag{36}$$

A comparison of the amplifying factor for the different equations at different grid optical thickness is presented in Figure 4. Here the single scattering albedo is set as 0.9. It is seen that the amplifying factor of discretization based on the RTE shows great values both near the low frequency limit and the high frequency limit. While the amplifying factor of discretization based on the MSORTE and the SORTE has low value at high frequency. With the increasing of grid optical thickness, the amplifying factor of each transfer equation tends to increase, which will result in slow convergence and is consistent with numerical experience. The amplifying factor of discretization based on the MSORTE and the SORTE is generally smaller than that based on the RTE for different frequencies. Furthermore, amplifying factor of discretization based on the MSORTE gives the lowest value at high frequency. This indicates that the discretization based on the MSORTE and the SORTE have better convergence performance than that based on the RTE, and the discretization based on the MSORTE has the best convergence performance.

The diffusion limit ($\beta \to \infty$) convergence behavior can be obtained through limit analysis of Eq. (34) (for the RTE), Eq. (36) (for the SORTE) and Eq. (35) (for the MSORTE). It is observed that the amplifying factors of discretization based on the RTE, the MSORTE and the SORTE all approach to $\omega$ and independent of frequency at the diffusion limit, which indicates these equations have similar convergence characteristics at diffusion limit and the convergence speed is determined only by the scattering albedo.

It is noted that there are some advanced techniques to improve the convergence speed of the source iteration, such as the DSA scheme [46, 47] and other acceleration methods [48]. These advanced techniques can also be applied to schemes based on the MSORTE to accelerate the convergence speed. The procedure is similar to the application of the DSA scheme to the SORTE [49].

## 4. Meshless Method Solution

The salient feature of the meshless methods is that it provides the ways of building approximation space just based on scattered nodes without relying on any mesh that is essential in traditional methods, such as FVM and FEM. The commonly used ways of building approximation space in meshless methods include moving least squares (MLS) approximation, reproducing kernel approximation and radial basis function approximation, of which the MLS approximation can be considered to be the most popular in the solution of PDEs [38]. In the following part, the MLS scheme is used to build the approximation space $U = \text{span}\{N_i(\mathbf{x}), i = 1, ..., N_{sol}\}$, where $N_i(\mathbf{x})$ is the nodal basis function of the $i$-th node. Then the approximation space is combined with the collocation approach to discretize the MSORTE.

### 4.1. Moving least square approximation

The basic idea of the MLS approximation is to *locally* project the unknown fields into a prescribed function space by least squares approach. Hence it will yield projections different for different positions. By tuning the function space, the MLS approximation can be very accurate. The most often used function space is the polynomial space. The local projection can be achieved through weight functions with compact support property. To give the local projection, we first define a local inner product as

$$<f,g>_{\bar{\mathbf{x}}} = \sum_{i=1}^{N_{sol}} f(\mathbf{x}_i)g(\mathbf{x}_i)w(|\mathbf{x}_i - \bar{\mathbf{x}}|) \tag{37}$$

where $w$ is a weight function with compact support, $\bar{\mathbf{x}}$ is the predefined position vector and $\mathbf{x}_i$ are the solution nodes of the computational domain. Though Eq. (37) is globally defined, it is meaningful only in the support domain of $w$ around $\bar{\mathbf{x}}$. The local inner product is thereafter used to define an error norm for the local minimization in the least squares approach.

The unknown intensity field $I(\mathbf{x})$ can be approximated locally around $\bar{\mathbf{x}}$ as

$$\tilde{I}_{\bar{\mathbf{x}}}(\mathbf{x}) = \sum_{l=0}^{P} p_l(\mathbf{x})\alpha_l(\bar{\mathbf{x}}) \tag{38}$$

where $p_l$ is the $l$-th polynomial basis function and $\alpha_l$ is the corresponding expansion coefficient. The least square approach can then be formulated with the help of the local inner product (around $\bar{\mathbf{x}}$) as

$$\min_{\alpha_l} J_{\bar{\mathbf{x}}}(\boldsymbol{\alpha}) = \min_{\alpha_l} \left\| \tilde{I}_{\bar{\mathbf{x}}} - I \right\|^2 = \min_{\alpha_l} < \sum_{l=0}^{P} p_l(\mathbf{x})\alpha_l - I(\mathbf{x}), \sum_{k=0}^{P} p_k(\mathbf{x})\alpha_k - I(\mathbf{x}) >_{\bar{\mathbf{x}}} \tag{39}$$

A solution to the minimization problem (using the extreme value condition $\partial J_{\bar{\mathbf{x}}}(\boldsymbol{\alpha})/\partial \alpha_l = 0$) yields

$$\sum_{l=0}^{P} < p_k(\mathbf{x}), p_l(\mathbf{x}) >_{\bar{\mathbf{x}}} \alpha_l = < p_k(\mathbf{x}), I(\mathbf{x}) >_{\bar{\mathbf{x}}} \tag{40}$$

which can be written in matrix form as

$$\boldsymbol{\Phi}\boldsymbol{\alpha} = \mathbf{f} \tag{41}$$

where

$$\boldsymbol{\Phi} = \left[\Phi_{kl}\right]_{k=1,P;l=1,P} = \left[< p_k(\mathbf{x}), p_l(\mathbf{x}) >_{\bar{\mathbf{x}}}\right]_{k=1,P;l=1,P} \tag{42a}$$

$$\boldsymbol{\alpha} = \left[\alpha_l\right]_{l=1,P}, \quad \mathbf{f} = \left[f_k\right]_{k=1,P} = \left[< p_k(\mathbf{x}), I(\mathbf{x}) >_{\bar{\mathbf{x}}}\right]_{k=1,P} \tag{42b}$$

The solution of the local expansion coefficients vector $\boldsymbol{\alpha}$ is then obtained as $\boldsymbol{\alpha} = \boldsymbol{\Phi}^{-1}\mathbf{f}$, and the locally approximated intensity field can be formulated as

$$\tilde{I}_{\bar{\mathbf{x}}}(\mathbf{x}) = \sum_{l=0}^{P} p_l(\mathbf{x})\alpha_l(\bar{\mathbf{x}}) = \mathbf{p}^T(\mathbf{x})\mathbf{\Phi}^{-1}(\bar{\mathbf{x}})\mathbf{f}(\bar{\mathbf{x}})$$
$$= \sum_{i=1}^{N_{sol}} \mathbf{p}^T(\mathbf{x})\mathbf{\Phi}^{-1}(\bar{\mathbf{x}})\mathbf{p}(\mathbf{x}_i)w(|\mathbf{x}_i - \bar{\mathbf{x}}|)I(\mathbf{x}_i)$$
(43)

where $\mathbf{p}(\mathbf{x}) = [p_l(\mathbf{x})]_{l=1,P}$. A nodal basis function (at node $i$) from the MLS approximation can then be obtained as

$$N_i(\mathbf{x};\bar{\mathbf{x}}) = \mathbf{p}^T(\mathbf{x})\mathbf{\Phi}^{-1}(\bar{\mathbf{x}})\mathbf{p}(\mathbf{x}_i)w(|\mathbf{x}_i - \bar{\mathbf{x}}|) \tag{44}$$

and the local polynomial approximation [Eq.(38)] can be formulated into the equivalent nodal basis expansion as

$$\tilde{I}_{\bar{\mathbf{x}}}(\mathbf{x}) = \sum_{i=1}^{N_{sol}} N_i(\mathbf{x};\bar{\mathbf{x}})I_i \tag{45}$$

where $I_i$ denotes $I(\mathbf{x}_i)$. By making the local coordinate moving, namely, setting $\bar{\mathbf{x}} = \mathbf{x}$ (i.e. moving with calculating point $\mathbf{x}$), the final MLS approximation is obtained as

$$\tilde{I}(\mathbf{x}) = \sum_{i=1}^{N_{sol}} N_i(\mathbf{x})I_i \tag{46}$$

where $N_i(\mathbf{x}) = N_i(\mathbf{x};\mathbf{x}) = \mathbf{p}^T(\mathbf{x})\mathbf{\Phi}^{-1}(\mathbf{x})\mathbf{p}(\mathbf{x}_i)w(|\mathbf{x}_i - \mathbf{x}|)$ is the MLS nodal basis function at node $i$. The explicit formulation of the $N_i(\mathbf{x})$ reveals that it is a function modulated by the weight function $w(|\mathbf{x}_i - \mathbf{x}|)$. Hence the selection of a compact support weight function will make the $N_i(\mathbf{x})$ possess the compact support property. A sparse, narrow-banded stiff matrix will be obtained with the compact support nodal basis function to discretize the PDEs.

During the solution of the MSORTE, first- and second-order partial derivatives of the nodal basis function are required. Here, the evaluation of the partial derivatives is by the 'diffuse derivative' approach [50], which can generally be written as

$$\nabla N_i(\mathbf{x}) = \nabla \mathbf{p}^T(\mathbf{x})\mathbf{\Phi}^{-1}(\mathbf{x})\mathbf{p}(\mathbf{x}_i)w(|\mathbf{x} - \mathbf{x}_i|) \tag{47a}$$

$$\nabla\nabla' N_i(\mathbf{x}) = \left[\nabla\nabla'\mathbf{p}^T(\mathbf{x})\right]\mathbf{\Phi}^{-1}(\mathbf{x})\mathbf{p}(\mathbf{x}_i)w(|\mathbf{x} - \mathbf{x}_i|) \tag{47b}$$

where $\nabla$ and $\nabla'$ can be considered as gradient operator or any first order partial differential operators. The second order partial derivatives are obtained in the same way. The diffuse derivative approach is based on the fact that the polynomial expansion [Eq. (38)] determined by the local least square approximation varies much slowly from point to point. Despite not an exact approach, it is reasonable to locally omit the variation of the expansion coefficients. The partial derivatives of the nodal basis functions can also be evaluated accurately by considering the variation of the expansion coefficients $\boldsymbol{\alpha}$, which is related to the part $\boldsymbol{\Phi}^{-1}(\mathbf{x})\mathbf{p}(\mathbf{x}_i)w(|\mathbf{x}-\mathbf{x}_i|)$ in Eq. (47) as shown in Eq. (43). However, numerical tests show that the exact evaluation can only give better results if the intensity distribution is smooth enough. If the intensity distribution contains discontinuity, the results will deteriorate. The focus of the present work is on radiative transfer in inhomogeneous media, which commonly may result in discontinuous intensity fields. In this condition, the diffuse derivative approach is a better choice. Otherwise, the exact evaluations of the partial derivatives are recommended.

In the following work, the function space spanned by the complete 2-D polynomial basis set up to second order, namely, $\{1, x, y, x^2, xy, y^2\}$, is used for the MLS approximation. To define the weight function $w(r)$, the 4-th order Wendland's function [51], namely,

$$\tilde{w}(r) = \left(1-r^2\right)^6 \left(35r^2 + 18r + 3\right), \quad r \leq 1 \tag{48}$$

is chosen as a template, which is compactly supported and continuous up to the 4-th order derivatives. Considering the radius of the support domain is $d$, then the weight function is defined as $w(r) = \tilde{w}(r/d)$. Generally, the determination of the support domain is related to local discrete length scale $\Delta h$ and the order $P$ of polynomial basis. A relation, $d = \alpha(P+1)\Delta h$, is recommended for determining the support domain of the weight function, and $\alpha = 1.5$ is used in this paper.

*4.2. Collocation approach discretization*

Theoretically, the collocation approach is a special form of the weighted residual approach with the weight function chosen as the Dirac delta function. Here, a collocation meshless method based on the MSORTE is presented to solve radiative transfer in inhomogeneous media. By substitution of the MLS nodal basis function approximation of the intensity field [Eq. (46)] and the approximation given below

$$\beta(\mathbf{x})I(\mathbf{x}) \approx \sum_{i=1}^{N_{sol}} N_i(\mathbf{x})\beta_i I_i \qquad S(\mathbf{x},\boldsymbol{\Omega}) \approx \sum_{i=1}^{N_{sol}} N_i(\mathbf{x})S_i(\boldsymbol{\Omega}) \tag{49}$$

into the MSORTE [Eq. (11)] yields

$$\sum_{i=1}^{K}\left\{(\mathbf{\Omega}\cdot\nabla)^2 N_i(\mathbf{x}) + \beta_i \mathbf{\Omega}\cdot\nabla N_i(\mathbf{x})\right\} I_i$$
$$= \sum_{i=1}^{K} \mathbf{\Omega}\cdot\nabla N_i(\mathbf{x}) S_i(\mathbf{\Omega}) \tag{50a}$$

which is formulated in the two-dimensional case as

$$\sum_{i=1}^{N_{sol}}\left\{\left[\mu^2 \frac{\partial^2}{\partial x^2} + 2\mu\eta \frac{\partial^2}{\partial x \partial y} + \eta^2 \frac{\partial^2}{\partial y^2}\right] N_i(\mathbf{x}) + \beta_i\left(\mu\frac{\partial}{\partial x} + \eta\frac{\partial}{\partial y}\right) N_i(\mathbf{x})\right\} I_i$$
$$= \sum_{i=1}^{N_{sol}}\left(\mu\frac{\partial}{\partial x} + \eta\frac{\partial}{\partial y}\right) N_i(\mathbf{x}) S_i \tag{50b}$$

and formulated in the one-dimensional case as

$$\sum_{i=1}^{K}\left\{\xi^2 \frac{d^2 N_i}{dz^2} + \beta_i \frac{dN_i}{dz}\right\} I_i = \sum_{i=1}^{K} \xi \frac{dN_i}{dz} S_i \tag{50c}$$

By using collocation approach, the discretized linear system can be obtained by requiring that Eq. (50) be exactly satisfied on each solution nodes with index $j$. The final discretized linear system for each discrete angular direction $\mathbf{\Omega}_m$ can be written in matrix form as

$$\mathbf{K}_m \mathbf{u}_m = \mathbf{h}_m \tag{51}$$

where subscript '$m$' denotes the index of discrete ordinates direction. The definitions of the matrices in Eq. (51) for two dimensional case are

$$\mathbf{u}_m = \left[u_{m,i}\right]_{i=1,N_{sol}} = \left[I_m(\mathbf{x}_i)\right]_{i=1,N_{sol}} \tag{52a}$$

$$\mathbf{K}_m = \mu_m^2 \mathbf{D}_{xx} + 2\mu_m \eta_m \mathbf{D}_{xy} + \eta_m^2 \mathbf{D}_{yy} + \left(\mu_m \mathbf{D}_x + \eta_m \mathbf{D}_y\right)\mathbf{B} \tag{52a}$$

$$\mathbf{h}_m = \left(\mu_m \mathbf{D}_x + \eta_m \mathbf{D}_y\right)\mathbf{s}_m \tag{52b}$$

in which the related matrices and vectors are defined as

$$\mathbf{B} = diag\{\mathbf{b}\} \tag{53a}$$

$$\mathbf{b} = \left[b_i\right] = \left[\beta(\mathbf{x}_i)\right] \tag{53b}$$

$$\mathbf{s}_m = \left[s_{m,i}\right] = \left[S(\mathbf{x}_i, \mathbf{\Omega}_m)\right] \tag{53c}$$

$$\mathbf{D}_x = \left[D_{x,ji}\right] = \left[\partial N_i(\mathbf{x}_j)/\partial x\right] \tag{53d}$$

$$\mathbf{D}_y = \left[ D_{y,ji} \right] = \left[ \partial N_i(\mathbf{x}_j) / \partial y \right] \tag{53e}$$

$$\mathbf{D}_{xx} = \left[ D_{xx,ji} \right] = \left[ \partial^2 N_i(\mathbf{x}_j) / \partial x^2 \right] \tag{53f}$$

$$\mathbf{D}_{xy} = \left[ D_{xy,ji} \right] = \left[ \partial^2 N_i(\mathbf{x}_j) / \partial x \partial y \right] \tag{53g}$$

$$\mathbf{D}_{yy} = \left[ D_{yy,ji} \right] = \left[ \partial^2 N_i(\mathbf{x}_j) / \partial y^2 \right] \tag{53h}$$

Here the matrix indices $i$ and $j$ all run from 1 to $N_{sol}$. For one dimensional case, the definitions of the matrices in Eq. (51) are given as follows.

$$\mathbf{K}_m = \xi_m^2 \mathbf{D}_{zz} + \xi_m \mathbf{D}_z \mathbf{B} \tag{54a}$$

$$\mathbf{h}_m = \xi_m \mathbf{D}_z \mathbf{s}_m \tag{54b}$$

in which, $\mathbf{D}_z$, $\mathbf{D}_{zz}$ are the one-dimensional first- and second-order discrete differentiation matrices defined similar to those defined in Eq. (53).

As for comparison, the collocation meshless method based on the RTE [Eq. (1)] and the SORTE [Eq.(6)] is also presented. The similar procedure is applied for discretization. For the RTE, the finally obtained matrices $\mathbf{K}_m$ and $\mathbf{h}_m$ in Eq. (51) for two dimensional case are

$$\mathbf{K}_m = \mu_m \mathbf{D}_x + \eta_m \mathbf{D}_y + \mathbf{B} \tag{55b}$$

$$\mathbf{h}_m = \mathbf{s}_m \tag{55c}$$

and for one dimensional case are

$$\mathbf{K}_m = \xi_m \mathbf{D}_z + \mathbf{B} \tag{56a}$$

$$\mathbf{h}_m = \mathbf{s}_m \tag{56b}$$

For the SORTE, only the formulation for media with constant extinction coefficient is presented, and the finally obtained matrices $\mathbf{K}_m$ and $\mathbf{h}_m$ in Eq. (51) for two dimensional case are

$$\mathbf{K}_m = -\left( \mu_m^2 \mathbf{D}_{xx} + 2\mu_m \eta_m \mathbf{D}_{xy} + \eta_m^2 \mathbf{D}_{yy} \right) + \beta^2 \tag{57b}$$

$$\mathbf{h}_m = \left[ \beta - \left( \mu_m \mathbf{D}_x + \eta_m \mathbf{D}_y \right) \right] \mathbf{s}_m \tag{55c}$$

and for one dimensional case are

$$\mathbf{K}_m = -\xi_m^2 \mathbf{D}_{zz} + \beta^2 \tag{58a}$$

$$\mathbf{h}_m = \left(\beta - \xi_m \mathbf{D}_z\right) \mathbf{s}_m \tag{56b}$$

It can be seen from Eqs. (53) that the tool matrices, namely, the series of discrete differentiation matrices $\mathbf{D}$ and $\mathbf{B}$, account for the major portion of the spatial discretization. The salient feature of these tool matrices is that they are independent of angular coordinates. Hence they only need to be calculated once for all angular loops and the source updating iterations for the discrete ordinates solution algorithm, which significantly reduces the computational effort in common solution process.

*4.3. Angular space quadrature*

Besides spatial discretization, angular space quadrature is another key factor that dictates the numerical solution accuracy of radiative heat transfer problems. The selection of a proper angular quadrature scheme in the discrete ordinates approach is often essential for an efficient solution. That is because too many angular directions are often needed to be solved to satisfy the accuracy of the heat flux calculation, which requires much computational effort and is mainly determined by the angular quadrature scheme when a space discretization has been chosen. For the inverse radiation transfer problems (such as the inverse geometrical design problem, to which the meshless method is a very promising approach), the computation effort is much more intensive, and the choosing of angular space quadrature scheme is more important.

A proper choice of solid angular quadrature scheme should explore the characteristics of the angular distribution of radiative intensity. The major characteristics include symmetry and smoothness. Generally, the solid angular space is two dimensional and described by the zenith angle $\theta$ and azimuthal angle $\varphi$ which is shown in Figure 5. For one-dimensional problems, the angular distribution of intensity is axisymmetrical (around axis defined in Figure 6). For two-dimensional problems, the intensity of positive *z* direction and negative *z* direction is symmetrical. These symmetry characteristics can be used to reduce discrete ordinate directions. As for smoothness, if the radiative intensity is estimated to distribute smooth enough with one angular coordinate, the known best way to integrate in that coordinate is by a Gaussian type quadrature (e.g. Gauss, Gauss-Legendre, Gauss-Radu and Gauss-Labotto, etc.), because which leads to fewest discrete directions for smooth functions, otherwise, a low order scheme, such as piecewise constant approach is appropriate. The design principle of the $S_N$ quadrature scheme is similar to Gaussian quadrature, and hence is suitable for quadrature of smooth angular distribution of intensity.

For the one-dimensional case, the radiative intensity is only a function of zenith angle $\theta$ (Figure 6) due to axisymmetry. Exploring the axisymmetry, the radiative transfer equation is simplified by integration with azimuthal angle and divided by $2\pi$. Furthermore, the radiative intensity often varies smoothly with $\theta$. Hence Gaussian type integration is very suitable in this case. The $S_N$ quadrature is a good choice, but currently the order of the $S_N$ can not be too high, which limited its applicability. In this paper, the standard Gaussian quadrature is used. The solid angular quadrature is carried out as follows

$$G(z) = 2\pi \int_0^\pi I(z,\theta) \sin\theta \mathrm{d}\theta = 2\pi \int_{-1}^1 I(z,\xi) \mathrm{d}\mu = 2\pi \sum_{m=1}^M I(z,\xi_m) \overline{w}_m \tag{59a}$$

and

$$q(z) \doteq 2\pi \sum_{m=1}^{N_\theta} I(z,\xi_m) \xi_m \overline{w}_m \tag{59b}$$

where $\mu_m$ is the *m*-th coordinate determined by Gaussian quadrature rule and $\overline{w}_m$ is the corresponding weight. It is noted that the scattering phase function should be modified accordingly to account for the simplification, generally, which takes the form

$$\tilde{\Phi}(\mu,\mu') = \frac{1}{2\pi} \int_0^{2\pi} \int_0^{2\pi} \Phi(\mathbf{\Omega},\mathbf{\Omega}') \mathrm{d}\varphi \mathrm{d}\varphi' \tag{60}$$

which is $2\pi$ and $2\pi(1+a\mu\mu')$ for isotropic ($\Phi=1$) phase function and linear anisotropic scattering ($\Phi=1+a\mathbf{\Omega}\cdot\mathbf{\Omega}'$) phase function, respectively.

For the two-dimensional case, the radiative intensity is symmetrical to *z* = 0 plane (Figure 5). By exploring the symmetry, only half of the total angular space is needed to be discretized. Similar to the one-dimensional case, the radiative intensity often varies smoothly with $\theta$ because of the uniform boundary condition. Hence Gaussian type integration is very suitable for this angular coordinates. However, the azimuthal distribution of radiative intensity is often discontinuous because of the boundary conditions and the block effect [8, 52]. As a result, a piecewise constant approach is suitable for the azimuthal direction quadrature. The solid angular integration can be calculated as

$$G(\mathbf{x}) = 2\int_0^{2\pi}\int_0^{\pi/2} I(\mathbf{x},\mathbf{\Omega})\sin\theta\,\mathrm{d}\theta\,\mathrm{d}\varphi = 2\int_0^{2\pi}\int_0^{1} I(\mathbf{x},\mathbf{\Omega})\,\mathrm{d}\mu\,\mathrm{d}\varphi = 2\sum_{m=1}^{M} I(\mathbf{x},\mathbf{\Omega}_m)\overline{w}_m \tag{61a}$$

and

$$\mathbf{q}(\mathbf{x}) = \sum_{m=1}^{M} I(\mathbf{x},\mathbf{\Omega}_m)\left[\mathbf{\Omega}_m + \overline{\mathbf{\Omega}}_m\right]\overline{w}_m \tag{61b}$$

where the discrete ordinate weight $\overline{w}_m = \overline{w}_i^{\theta}\overline{w}_j^{\varphi}$, in which $m = i+(j-1)N_\theta$, $i=1,...,N_\theta$, $j=1,...,N_\varphi$, $\overline{w}_i^{\theta}$ are the weight determined by Gaussian quadrature rule for zenith coordinate and $\overline{w}_j^{\varphi}$ is the weight determined by the piecewise constant approach for azimuthal coordinate, $\mathbf{\Omega}_m = \mathbf{\Omega}(\theta_i,\varphi_j)$ is the corresponding direction, $\overline{\mathbf{\Omega}}_m = [\mu_m,\eta_m,-\xi_m]$ is the mirrored direction of $\mathbf{\Omega}_m$ to the symmetrical plane. The scattering phase function also needs to be modified accordingly to account for the simplification, which takes the form

$$\tilde{\Phi}(\mathbf{\Omega},\mathbf{\Omega}') = \Phi(\mathbf{\Omega},\mathbf{\Omega}') + \Phi(\mathbf{\Omega},\overline{\mathbf{\Omega}}') \tag{62}$$

The quadrature for the modified phase function is carried over the $2\pi$ angular space.

It is noted that there are some other principles on the design of angular quadrature schemes, such as requiring the quadrature scheme to be conservative such that the scattering integral being accurate [53, 54].

*4.4. Boundary condition treatment and implementation*

There are two kinds of boundary condition for the MSORTE (same as the SORTE). The inflow boundary condition is of the Dirichlet type, and the outflow boundary condition is of the Neumann type. Here both the boundary conditions are imposed by operator collocation, namely, replacing the row of stiff matrix $\mathbf{K}_m$ corresponding to the boundary nodes with the discrete operator of the related boundary condition. Similar modification are also applied to the load vector $\mathbf{h}_m$. The modification process of the stiff matrix $\mathbf{K}^m$ and load vector $\mathbf{h}_m$ are formulated as follows

$$K_{m,ji} = \delta_{ji}; \quad h_{m,j} = I_0(\mathbf{x}_j, \mathbf{\Omega}_m), \quad \mathbf{n}_w(\mathbf{x}_j) \cdot \mathbf{\Omega}_m < 0 \tag{63a}$$

$$K_{m,ji} = \mu_m D_{x,ji} + \eta_m D_{y,ji} + B_{ji}; \quad h_{m,j} = s_{m,j}, \quad \mathbf{n}_w(\mathbf{x}_j) \cdot \mathbf{\Omega}_m \geq 0 \tag{63b}$$

for two dimensional case or

$$K_{m,1i} = \delta_{1i}; \quad h_{m,1} = I_0(0, \xi_m), \quad \xi_m > 0 \tag{64a}$$

$$K_{m,1i} = \xi_m D_{z,1i} + B_{1i}; \quad h_{m,j} = s_{m,j}, \quad \xi_m > 0 \tag{63b}$$

$$K_{m,N_{sol}i} = \delta_{N_{sol}i}; \quad h_{m,N_{sol}} = I_0(L, \xi_m), \quad \xi_m < 0 \tag{65c}$$

$$K_{m,N_{sol}i} = \xi_m D_{z,N_{sol}i} + B_{N_{sol}i}; \quad h_{m,N_{sol}} = s_{m,N_{sol}}, \quad \xi_m < 0 \tag{63d}$$

for one dimensional case.

The solution procedures of the meshless method are similar to other discrete ordinates based methods, which comprise an outer iteration of global source update and an inner angular loop. The implementation of the presented meshless method is carried out according to the procedures depicted in Figure 7.

## 5. Numerical Experiment

Several critical test cases of one- and two- dimension are selected to fully verify the performance of the meshless methods based on the MSORTE. The maximum relative error of the incident radiation $\|G_{new} - G_{old}\| / \|G_{new}\| < 10^{-4}$ is taken as stopping criterion for the global source update iteration.

### 5.1. One dimensional test cases

#### 5.1.1. Gaussian shaped emissive field in an infinite slab

The radiative transfer in an absorbing-emitting infinite slab is considered. The extinction coefficient of the medium is a constant, but the emissive field has a Gaussian profile. This problem is modeled by the RTE as

$$\xi \frac{dI}{dz} + \kappa_a I = e^{-(z-c)^2/\alpha^2}, \quad z, c \in [0,1] \tag{66}$$

with boundary condition

$$I(0, \mu) = \kappa_a^{-1} e^{-c^2/\alpha^2}, \quad \mu > 0 \tag{67}$$

in which $c$ is a parameter to denote the location of the peak of the Gaussian profile and $\alpha$ is a parameter to change the shape of the Gaussian profile. The analytical solution of this problem for $\mu > 0$ is

$$\begin{aligned} I(z, \mu) = & I(0, \mu) \exp\left(-\frac{\kappa_a z}{\mu}\right) \\ & - \frac{\alpha \sqrt{\pi}}{2\mu} \exp\left\{-\frac{\kappa_a}{\mu}\left[z - \left(\frac{\alpha^2 \kappa_a}{4\mu} + c\right)\right]\right\} \left[\text{erf}\left(\frac{\alpha \kappa_a}{2\mu} + \frac{c-z}{\alpha}\right) - \text{erf}\left(\frac{\alpha \kappa_a}{2\mu} + \frac{c}{\alpha}\right)\right] \end{aligned} \tag{68}$$

This problem features large gradient of intensity distribution, which was first proposed to test the performance of the SORTE by using FEM [33]. Here, the meshless methods based on the MSORTE are applied to this case for $c = 0.5$, $\alpha = 0.02$, and three different optical thickness namely 0.1, 1 and 10. The slab is uniformly subdivided into 61 nodes. The solved distribution of radiative intensity in the direction $\mu = 0.5$ is presented in Figure 8(a) and compared with the exact analytical solutions. The results obtained by the meshless method based on the RTE under the same condition presented in Figure 8(b). As seen from the figures, the results obtained by the meshless method based on the RTE shows enormous spurious oscillations, while the results obtained by the meshless method based on the SORTE and the MSORTE are accurate and stable for all three values of optical thickness. This case well demonstrates the good numerical properties of the MSORTE. Also, slightly accuracy improvement can be observed for the solution based on the MSORTE than that based on the SORTE near the peak of intensity distribution.

*5.1.2. Infinite slab with an inclusion layer*

The previous case demonstrates the stability and accuracy of the MSORTE in solving radiative transfer in a homogeneous slab. In this case, radiative transfer in an inhomogeneous slab with a non-scattering inclusion layer is studied, to further investigate their performance in dealing with inhomogeneous medium. The configuration of the slab is shown in Figure 9. The thickness of the slab is $L$ and the thickness of the layer is $d = 0.4L$. The optical thickness of the inclusion layer is defined as $\tau_d = \beta_I d$, where $\beta_I$ is the extinction coefficient of the inclusion layer. The other portion of the slab is a void. The temperature of the left wall is 1000K, the temperature of the inclusion layer is $T_g = 500K$, and the right wall is kept cold. Both the left and the right walls are black.

The meshless methods based on the MSORTE are applied to solve this problem to demonstrate its capability and performance in dealing with a zero extinction coefficient. The slab is uniformly subdivided into 200 subdivisions (with 201 nodes), and the angular discretization using the Gauss integration with 20 discrete directions. The heat flux and incident radiation distribution in the slab produced by the meshless methods based on the MSORTE are shown in Figure 10(a) and (b), respectively. Here, the converged results obtained using the discontinuous spectral element method [52] are taken as the reference (exact) results. The results obtained by a meshless method based on the RTE under the same computational condition are shown in Figure 11.

It is seen that the heat flux and incident radiation distribution obtained by the meshless method based on the MSORTE agree very well with the reference results. However, the results obtained by the meshless method based on the RTE are totally spoiled by the spurious oscillations, which are caused by the steep gradient of intensity distribution. A function with steep gradient in frequency domain will have high frequency harmonic component, which will induce large numerical errors for the center difference discretization of the RTE as indicated by the theoretical analysis presented in Section 3. Furthermore, the strong wiggles with a wavelength of about double of the grid spacing appear in the solved intensity distribution agree well with the theoretical predication.

A spatial convergence test is conducted for the meshless method based on the MSORTE and compare with that based on the RTE for solving the incident radiation under $\tau_d$ = 10 and 1, which is presented in Figure 12. The relative error is defined based on $l_2$ norm as $\|G_{solved} - G_{exact}\| / \|G_{exact}\|$. It is seen both the methods show larger errors for a large value of inclusion optical thickness ($\tau_d = 10$). Furthermore, the former shows much lower level of relative error, i.e., smaller than the latter by nearly two orders magnitude, and converges faster than the latter.

This case demonstrates the good numerical stability and accuracy of the meshless method based on the MSORTE in solving radiative transfer in strong inhomogeneous media with a discontinuous extinction coefficient.

*5.2. Two dimensional test cases*

*5.2.1. Square enclosure with a Gaussian shaped emissive field*

The radiative transfer in an absorbing-emitting square enclosure is considered. The emissive field of the medium has a Gaussian profile. This test case is an extension of the one-dimensional case studied in Section 5.1.1 to two dimensions, which is taken here to especially demonstrate the stability of the presented methods in multi-dimensions. This problem is modeled by the RTE as

$$\mu \frac{dI}{dx} + \eta \frac{dI}{dy} + \kappa_a I = e^{-[\tilde{s}(x,y)-c]^2/\alpha^2}, \quad x, y \in [0,1] \tag{69}$$

in which the incident direction is selected as $\mu = \eta = \sqrt{2}/2$, $\tilde{s}(x,y) = (x+y)/\sqrt{2}$ is a distance parameter in the incident direction, the emission profile parameters are selected as $c = \sqrt{2}/2$ and $\alpha = 0.02$. The boundary condition is prescribed on the inflow boundary with null emission, namely, on the left and bottom wall for the considered direction. An analytical solution can be obtained based on an analysis along the characteristic lines as

$$I(x,y) = \frac{\alpha \sqrt{\pi}}{2} \exp\left\{-\kappa_a \left[s(x,y) - c_s(x,y) - \alpha^2 \kappa_a / 4\right]\right\} \\ \times \left[\mathrm{erf}\left(\frac{\alpha \kappa_a}{2} + \frac{c_s(x,y)}{\alpha}\right) - \mathrm{erf}\left(\frac{\alpha \kappa_a}{2} + \frac{c_s(x,y) - s(x,y)}{\alpha}\right)\right] \tag{70}$$

in which $s(x,y) = \dfrac{|x+y| - |x-y|}{\sqrt{2}}$ and $c_s(x,y) = \dfrac{1-|x-y|}{\sqrt{2}}$. Considering the exact solution of this problem has almost zero values in some location and to avoid overflow, the relative error $E(\mathbf{x})$ and averaged relative error $E_{\text{AVG}}$ in the solution domain are defined as

$$E(\mathbf{x}) = \frac{I_{\text{N}}(\mathbf{x}) - I_{\text{E}}(\mathbf{x})}{\max_{\mathbf{x}}[I_{\text{E}}(\mathbf{x})]} \tag{71}$$

$$E_{\text{AVG}} = \frac{1}{N_{sol}} \sum_{i=1}^{N_{sol}} |E(\mathbf{x}_i)| = \frac{1}{N_{sol}} \sum_{i=1}^{N_{sol}} \left| \frac{I_{\text{N}}(\mathbf{x}_i) - I_{\text{E}}(\mathbf{x}_i)}{\max_{i}[I_{\text{E}}(\mathbf{x}_i)]} \right| \tag{72}$$

where $I_{\text{E}}$ denotes the exact result [given by Eq. (70)], $I_{\text{N}}$ denotes the numerical result.

The meshless methods based on the RTE, the MSORTE are applied to solve the intensity distribution in the medium for different value of $\kappa_a$ (=1, 10, 20 and 100), and the corresponding numerical error distributions are presented in Figure 13(a), (b) and (c), respectively. The square enclosure is subdivided uniformly into 421 nodes with a grid size of $\Delta x = \Delta y = 1/20$. In this condition, the grid optical thickness $\tau_\Delta = \kappa_a \Delta x$ for the selected absorption coefficient is 0.05, 0.5, 1 and 5, respectively. At the low grid optical thickness, as seen from Figure 13(a), the spurious oscillations dominate the result and the intensity distribution obtained using meshless method based on the RTE is almost totally spoiled. With the increasing of absorption coefficient, the averaged relative error tends to decrease, such as, decrease from $E_{\text{AVG}} = 0.768$ at $\kappa_a$ =1 to $E_{\text{AVG}} = 0.025$ at $\kappa_a$ =100. From Figure 13(b) and (c), it is seen that the numerical error distribution obtained based on the MSORTE are quantitatively similar. At $\kappa_a$ = 1, the $E_{\text{AVG}}$ is of about 0.04, which is about 20 times lower than that of the numerical error distribution obtained based on the RTE. This indicates that the MSORTE and the SORTE are stable at low grid optical thickness. With the increasing of grid optical thickness, the trend of the variation of the averaged relative error obtained based on the MSORTE and the SORTE does not decrease monotonically.

To get an in depth understanding of the trend of the variation of the averaged relative error with grid optical thickness, more computation is conducted (the maximum grid optical thickness considered is up to 25 with $\kappa_a$ = 500) and the averaged relative error obtained based on different equations versus $\kappa_a$ (and $\tau_\Delta$ ) is plotted in Figure 14(a). The theoretical error relation (at $\bar{\varpi} = 0.4$) with grid optical thickness from the frequency domain analysis derived in Section 3.1 is presented in Figure 14(b). Though the meshless discretization presented is similar to central difference discretization in numerical property, they are not exactly equivalent. Hence the theoretical error relation is only for a qualitative reference. From the numerical error relations for different equations obtained by the meshless method [Figure 14(a)], it is observed that there are three major characteristics of the relations: (1) a cross point (at about $\tau_\Delta = 5$) exists between the RTE relation and the MSORTE relation. When the grid optical thickness is greater than the cross point grid optical thickness, the performance of the RTE tends to be better than the MSORTE; (2) The SORTE error curve approaches the RTE error curve at large grid optical thickness, and the error decreases with the increase of grid optical thickness; (3) The error curves of the MSORTE and the SORTE coincide at low grid optical thickness. It is seen that the three major characteristics agree qualitatively well with the theoretically predicted trends. The good numerical stability of the MSORTE in solving multi-dimensional problems at low and medium optical thickness is demonstrated.

*5.2.2. Square enclosure with a homogeneous anisotropic scattering inclusion*

Radiative heat transfer in a square enclosure with a scattering inclusion is considered. The configuration of the square enclosure is depicted in Figure 15. The scattering inclusion is of square shape and centered in the square enclosure. The side length of the square enclosure is *L*, and the side length of the square inclusion is *d* = 0.5*L*. The medium in the enclosure is non-participating except the region of the scattering inclusion. The temperature of all the walls is 500*K*, while the temperature of the square inclusion is 1000*K*. The walls of the enclosure are all black. The optical thickness of the square inclusion is defined based on its side length set as $\tau_I = \beta_I d = 1.0$. The anisotropic scattering phase function is given as [55]

$$\Phi(\mathbf{\Omega},\mathbf{\Omega}') = \sum_{l=0}^{8} C_l P_l(\mathbf{\Omega}\cdot\mathbf{\Omega}') \tag{73}$$

where $P_l$ is the *l*-th order Legendre polynomial and $C_l$ is the corresponding expansion coefficient given as

$C_0$=1.0, $C_1$=2.00917, $C_2$=1.56339, $C_3$=0.67407, $C_4$=0.22215, $C_5$=0.04725, $C_6$=0.00671, $C_7$=0.00068 and $C_8$=0.00005, respectively.

The meshless methods based on the RTE and the MSORTE are applied to solve the heat flux and incident radiation distribution ($G(\mathbf{x}) = \int_{4\pi} I(\mathbf{x},\mathbf{\Omega})\mathrm{d}\Omega$) for three different single scattering albedo of the inclusion, namely, $\omega_I = 0$, 0.5 and 0.9, under the same computational condition. The square enclosure is uniformly subdivided into 961 nodes, and the angular space is discretized into $N_\theta \times N_\varphi = 18 \times 72$ directions. The solved dimensionless net radiative heat flux distributions $q_w/\sigma T_{g,I}^4$ along the bottom wall through different meshless methods are presented in Figure 16. The solved incident radiation field distributions (represented as a false temperature as $T^* = \sqrt[4]{G/4\sigma}$) by different methods for $\omega_I = 0.5$ are presented in Figure 17, in which, Figure 17(a) gives the result obtained based on the RTE, and Figure 17(b) gives the results obtained based on the MSORTE. Here, the results obtained using the discontinuous spectral element method [52] are taken as the reference results.

It is observed that the heat flux distribution and the incident radiation field distribution obtained by the meshless method based on the RTE are full of nonphysical oscillations which are caused by the instability of the RTE discretized with central difference like method in dealing with very steep intensity field. The false temperature distribution obtained based on the RTE is almost totally spoiled by the spurious wiggles, hence it is plotted separately in Figure 17(a). However, the heat flux distributions and the incident radiation field distributions obtained by the meshless method based on the MSORTE are free of spurious oscillations and shows very good accuracy (such as in heat flux distribution, it gives a maximum relative error of less than 2% for $\omega_I = 0$). The excellent numerical stability of the MSORTE is attributed to the second order differential (diffusion) term. This case demonstrate the meshless method based on the MSORTE is capable of stably and accurately solve radiative transfer in multidimensional media with void region and discontinuous extinction coefficient.

*5.2.3. Semicircular enclosure with an inhomogeneous emitting-absorbing inclusion*

In this case, radiative heat transfer in a semicircular enclosure with an inhomogeneous emitting-absorbing inclusion is considered. The configuration of the semicircular enclosure is depicted in Figure 18(a). The absorption coefficient of the medium in the enclosure is a function of radial coordinates and defined as

$$\kappa_a(x,y) = \kappa_{a,I}\tilde{w}\left(\frac{2}{d}\sqrt{x^2 + (y-y_0)^2}\right) \qquad (74)$$

in which $\kappa_{a,I}$ is the maximum value of $\kappa_a(x,y)$, $d$ is a prescribed scale parameter and $\tilde{w}(\bullet)$ is a compactly supported radial basis function defined previously by Eq. (48). The definition of $\kappa_a(r)$ by Eq. (74) indicates that the absorption coefficient is non zero only for $\sqrt{x^2 + (y-y_0)^2} \leq d/2$, which defines an inhomogeneous inclusion in that region. The optical thickness of the inhomogeneous inclusion is defined as $\tau_I = \kappa_{a,I} d$. The overall medium is non-scattering with a temperature of $T_g = 1000K$. The temperature of the walls are kept $T_w = 500K$. In the following study, the translation parameter $y_0$ is taken as $y_0 = D/4$, the prescribed scale parameter $d$ is set as $d = 0.4D$, $D$ is the diameter of the enclosure.

The meshless methods based on the RTE and the MSORTE are applied to solve the heat flux and incident radiation distribution for three different inclusion optical thickness, namely, $\tau_I$ = 0.1, 1, and 2, under the same computational condition. The semicircular enclosure is subdivided into 1899 nodes shown in Figure 18(b), and the angular space is discretized into $N_\theta \times N_\varphi = 14 \times 56$ directions. The solved dimensionless net radiative heat flux distributions $q_w/\sigma T_g^4$ by different methods along the bottom wall are presented in Figure 19. The solved incident radiation field distributions (represented as a false temperature as $T^* = \sqrt[4]{G/4\sigma}$) by different methods for $\tau_I$ = 1 are presented in Figure 20. The results obtained using the discontinuous spectral element method [52] are taken as the reference results.

Again, spurious wiggles are observed in the solved the heat flux and incident radiation distribution by the methods based on the RTE, which induce very large errors near the boundary (with a maximum relative error of heat flux being greater than 100% for $\tau_I = 2$). The results obtained based on the MSORTE gives good accuracy (with a maximum relative error of less than 2.5% for $\tau_I = 2$). This further demonstrates the good numerical stability and accuracy of the meshless method based on the MSORTE for strongly inhomogeneous medium.

## 6. Conclusions

A new second order form of radiative transfer equation (MSORTE) is proposed. In the MSORTE, the second order partial differential (or diffusion) terms are naturally introduced and physically consistent with the original first order equation. The new second order radiative transfer equations overcome the singularity problem of the SORTE in dealing with inhomogeneous media where some locations have very small extinction coefficient.

The numerical stability and convergence characteristics of the MSORTE discretized by the central difference scheme are studied in an analytical framework in frequency domain and compared with that of the RTE and the SORTE. It is proved that the intensity distribution with large gradient will induce the instability problem (namely, strong spurious wiggles with wavelength about two times the grid spacing appear in the solved intensity distribution) when solving using the central difference type scheme based on the RTE. However, the central difference type method based on the second order equations owns very good stability.

A collocation meshless method based on the MSORTE is developed and demonstrated to be numerically stable and accurate in solving radiative transfer in strongly inhomogeneous media, media with void region and even with discontinuous extinction coefficient.

It is noted that the theoretical analysis of the performance of the MSORTE presented in this work is only for media with homogeneous extinction coefficient. By experience, the evaluation based on the analysis for homogeneous media may not be still valid for inhomogeneous media. Furthermore, since the meshless method is based upon a trial space with continuous derivatives, it is still suspect for application to any form of the transport equation for problems with discontinuity in extinction coefficient. Though some success is attained for the meshless method based on the MSORTE for solving radiative transfer in strongly inhomogeneous media, the general robustness of the central difference type method based on the MSORTE needs to be examined further. And the evaluation of the performance of other numerical methods based on the MSORTE is also appealing.

**Acknowledgements**

The support of this work by the National Nature Science Foundation of China (50906017, 50836002) and the Development Program for Outstanding Young Teachers in Harbin Institute of Technology (HITQNJS.2009.020) are gratefully acknowledged. J.M. Zhao appreciates Professor Denis Lemonnier very much for the valuable discussions with him on the SORTE during the period of the *6$^{th}$ International Symposium on Radiative Transfer* (*Rad-10*). The valuable advices from the three reviewers are also acknowledged very much.

**Appendix**

**A. Frequency domain relations of the finite difference discretization**

By using Taylor expansion, intensity at current analyzed node (node *n*) can be used to express the intensity at neighboring nodes, such as

$$I_{n+1} = I_n + \sum_{k=1}^{\infty} (\Delta s)^k \frac{1}{k!} \frac{d^k I_n}{ds^k} \tag{75a}$$

$$I_{n-1} = I_n + \sum_{k=1}^{\infty} (-\Delta s)^k \frac{1}{k!} \frac{\partial^k I_n}{\partial s^k} \tag{75b}$$

The right hand side of Eq. (75) can be considered a differential formulation evaluated at coordinates $s = s_n$.

As such, a Fourier transform can be defined for $I_{n+1}$ and $I_{n-1}$ by taken $s_n$ as the corresponding time domain variable, which yield

$$\Im[I_{n+1}] = \widehat{I_{n+1}} = \sum_{k=0}^{\infty} \frac{1}{k!}(j\varpi\Delta s)^k \widehat{I_n} = e^{j\varpi\Delta s} \widehat{I_n} \tag{76}$$

$$\Im[I_{n-1}] = \widehat{I_{n-1}} = \sum_{k=0}^{\infty} \frac{1}{k!}(-j\varpi\Delta s)^k \widehat{I_n} = e^{-j\varpi\Delta s} \widehat{I_n} \tag{77}$$

In which both $\Im[\bullet]$ and the over triangle '^' denotes the Fourier transform operator.

In this principle, the central difference discretization of $dI/ds$ can be written in differential form as

$$\frac{I_{n+1} - I_{n-1}}{2\Delta s} = \frac{dI_n}{ds} + \frac{1}{2\Delta s}\sum_{k=2}^{\infty}\left[1-(-1)^k\right](\Delta s)^k \frac{1}{k!}\frac{d^k I_n}{ds^k} \tag{78}$$

and its Fourier transform can be obtained as

$$\Im\left[\frac{I_{n+1} - I_{n-1}}{2\Delta s}\right] = \frac{e^{j\varpi\Delta s}\widehat{I_n} - e^{-j\varpi\Delta s}\widehat{I_n}}{2\Delta s} = j\varpi\text{sinc}(\varpi\Delta s)\widehat{I_n} \tag{79}$$

Similarly, the central difference discretization of $d^2 I/ds^2$ can be written in differential form as

$$\frac{I^{n+1} - 2I^n + I^{n-1}}{(\Delta s)^2} = \frac{d^2 I^n}{ds^2} + \frac{1}{(\Delta s)^2}\left\{\sum_{k=3}^{\infty}\left[(-1)^k + 1\right]\Delta s^k \frac{1}{k!}\frac{d^k I^n}{ds^k}\right\} \tag{80}$$

and its Fourier transform can be obtained as

$$\Im\left[\frac{I^{n+1} - 2I^n + I^{n-1}}{(\Delta s)^2}\right] = \frac{e^{j\varpi\Delta s}\widehat{I_n} - 2\widehat{I_n} + e^{-j\varpi\Delta s}\widehat{I_n}}{(\Delta s)^2} = -\varpi^2\text{sinc}^2(\frac{\varpi\Delta s}{2})\widehat{I_n} \tag{81}$$

**Figure Captions**

**Figure 1.** Schematic of the definition of inflow and outflow boundary conditions for the MSORTE.

**Figure 2.** The frequency domain distribution of solution error for the central difference discretization based on the RTE, the MSORTE and the SORTE at different grid optical thickness.

**Figure 3.** Detailed relation of the solution error with grid optical thickness for the central difference discretization based on the RTE, the MSORTE and the SORTE at three different reduced frequencies, $\bar{\varpi} = 0.1$, 0.25 and 0.5.

**Figure 4.** The amplifying factors of source iteration for the central difference discretization based on the RTE, the MSORTE and the SORTE at different grid optical thickness.

**Figure 5.** Spatial and angular coordinates definition for the two-dimensional radiative transfer problem.

**Figure 6.** Schematic of the one-dimensional infinite slab and variable definition.

**Figure 7.** Flow chart of the implementation of the meshless methods.

**Figure 8.** Solved radiative intensity distribution in the infinite slab for direction $\mu = 0.5$ by meshless methods based on the first- and second- order equations: **(a)** on the second order equations and **(b)** on the first order equation.

**Figure 9.** Configuration of the inhomogeneous slab with an inclusion layer.

**Figure 10.** Solved radiative heat flux distribution **(a)** and incident radiation distribution **(b)** in the inhomogeneous slab by meshless methods based on the second order equations (gray vertical lines denote the inclusion interfaces).

**Figure 11.** Solved radiative heat flux distribution **(a)** and incident radiation distribution **(b)** in the inhomogeneous slab for different inclusion optical thickness by meshless methods based on the RTE (gray vertical lines denote the inclusion interfaces).

**Figure 12.** Spatial convergence test of the meshless method based on the MSORTE and the RTE for solving the incident radiation under different inclusion optical thickness.

**Figure 13.** The relative error distribution of intensity in the square enclosure solved by meshless method based on different equations: **(a)** RTE, **(b)** SORTE, **(c)** MSORTE.

**Figure 14.** The variation of the averaged relative error with grid optical thickness for the RTE, the MSORTE and the SORTE: **(a)** numerical results, **(b)** theoretical predictions.

**Figure 15.** Configuration of the square enclosure with a homogeneous anisotropic scattering inclusion.

**Figure 16.** Solved dimensionless net radiative heat flux distributions along the bottom wall of the square enclosure for different inclusion scattering albedo by meshless methods based on the RTE and the MSORTE.

**Figure 17.** Solved incident radiation field distributions (represented as a false temperature as $T^* = \sqrt[4]{G/4\sigma}$) for $\omega_I = 0.5$ in the square enclosure by meshless methods based on different equations: **(a)** the RTE, **(b)** the MSORTE (blue thick solid) and the reference result (red dashed) for comparison.

**Figure 18.** Configuration of the semicircular enclosure and solution nodes: **(a)** configuration and **(b)** 1899 nodes distribution.

**Figure 19.** Solved dimensionless net radiative heat flux distributions $q_w/\sigma T_g^4$ along the bottom wall of the semicircular enclosure for different inclusion optical thickness by meshless methods based on the RTE and the MSORTE..

**Figure 20.** Solved incident radiation field distributions (represented as a false temperature as $T^* = \sqrt[4]{G/4\sigma}$) for $\tau_I = 1$ in the semicircular enclosure by meshless methods based on the RTE (black thin solid), the MSORTE (blue thick solid) and the reference result (red dashed) for comparison.

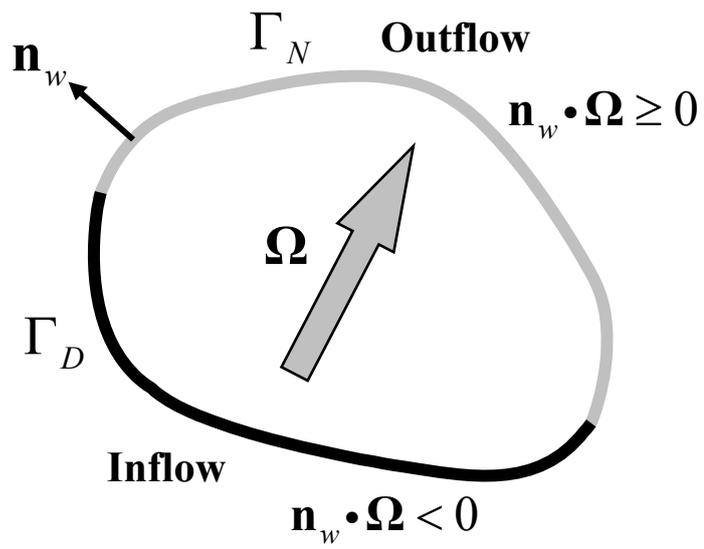

**Figure 1**

**Authors: Zhao, Tan and Liu**

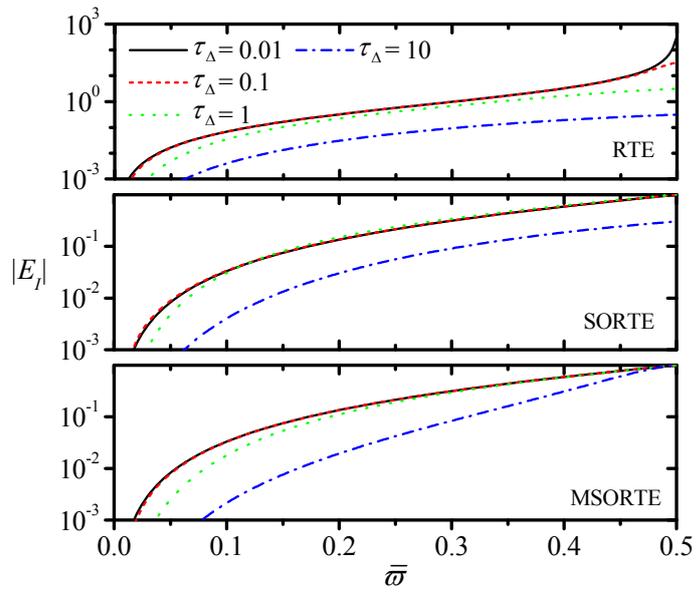

**Figure 2**

**Authors: Zhao, Tan and Liu**

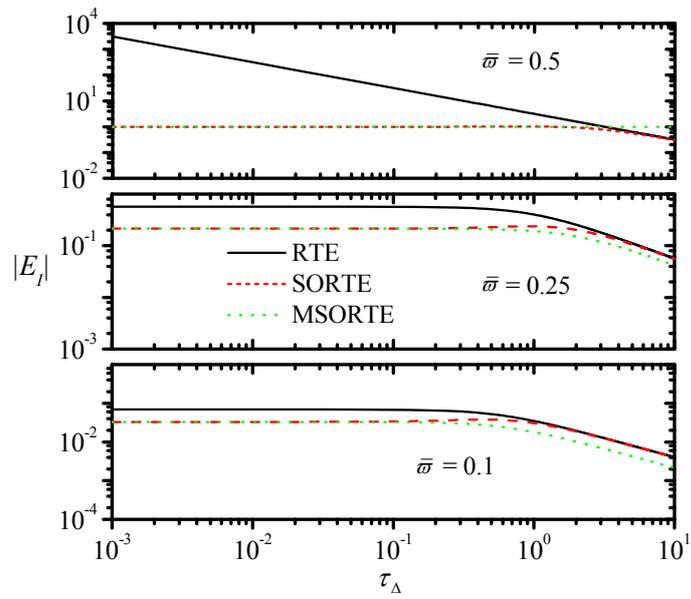

**Figure 3**

Authors: Zhao, Tan and Liu

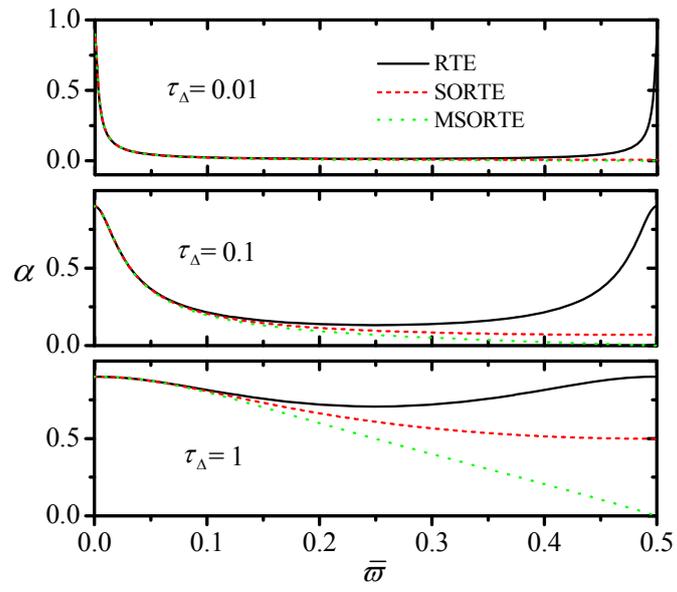

**Figure 4**

**Authors: Zhao, Tan and Liu**

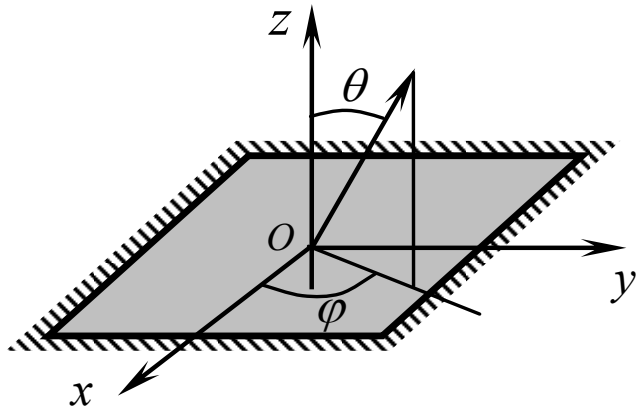

**Figure 5**

**Authors: Zhao, Tan and Liu**

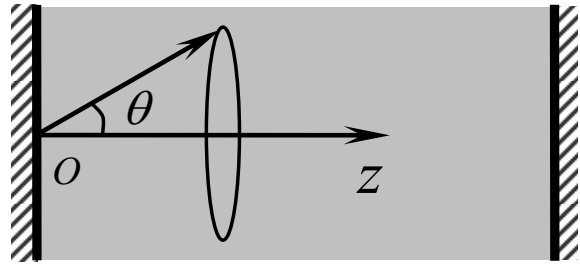

**Figure 6**

**Authors: Zhao, Tan and Liu**

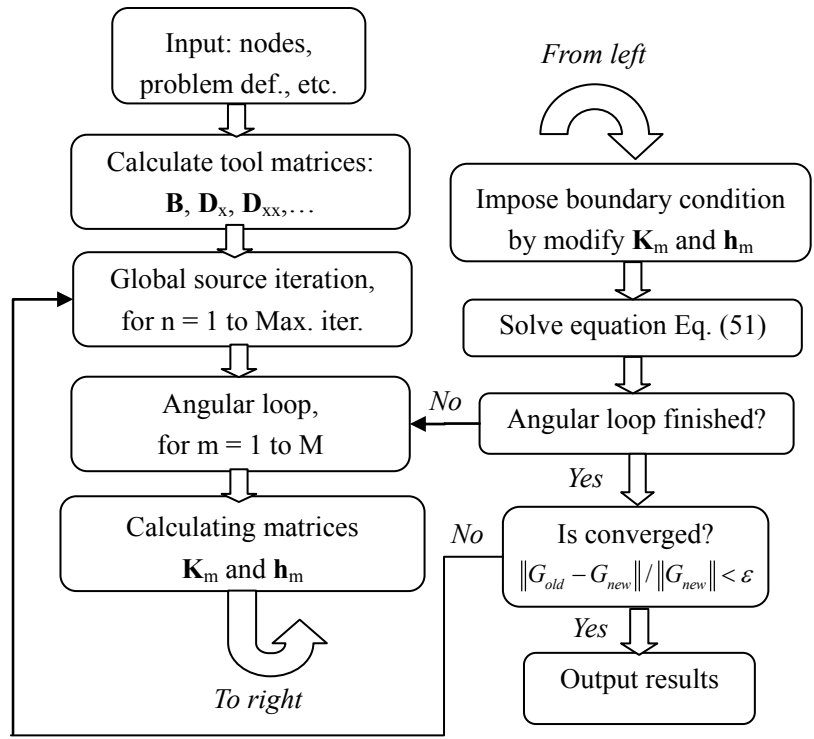

**Figure 7**

**Authors: Zhao, Tan and Liu**

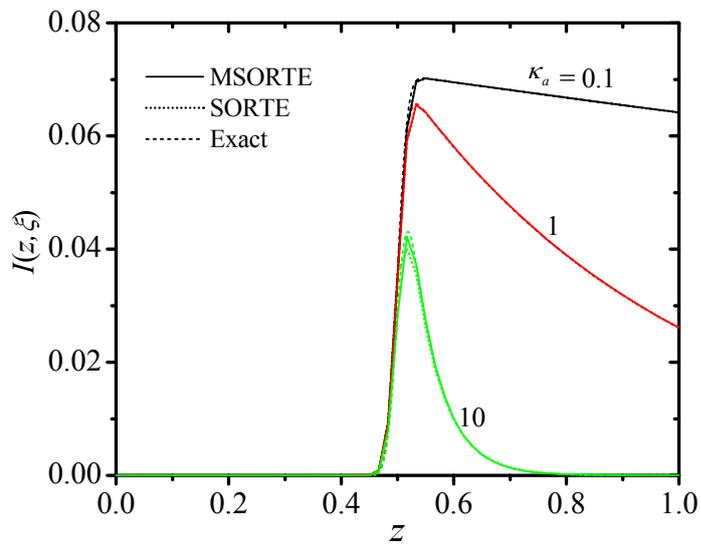

**Figure 8(a)**

**Authors: Zhao, Tan and Liu**

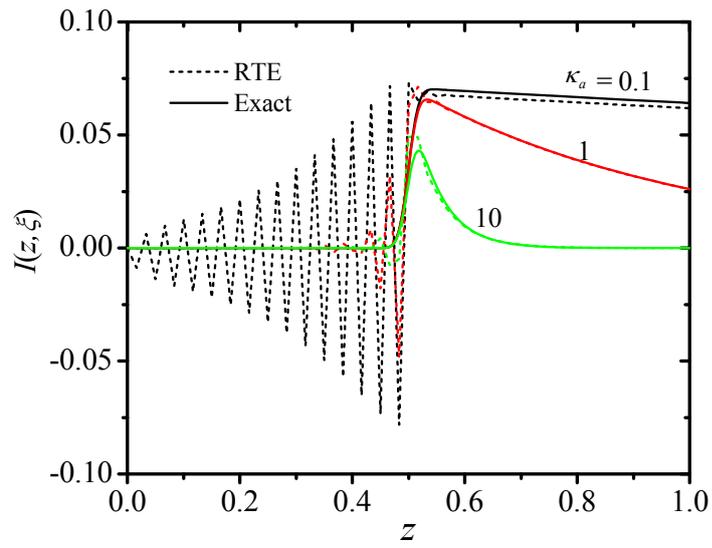

**Figure 8(b)**

**Authors: Zhao, Tan and Liu**

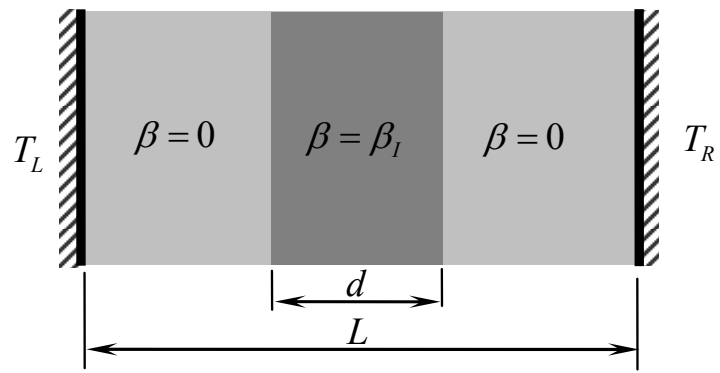

**Figure 9**

**Authors: Zhao, Tan and Liu**

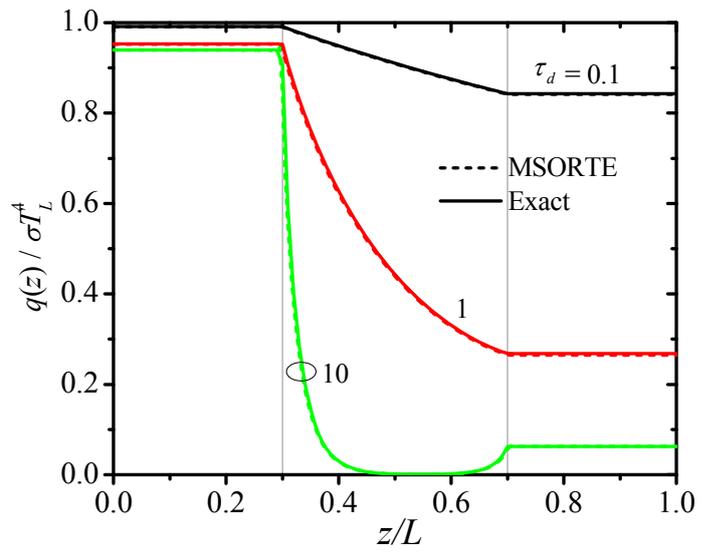

**Figure 10(a)**

**Authors: Zhao, Tan and Liu**

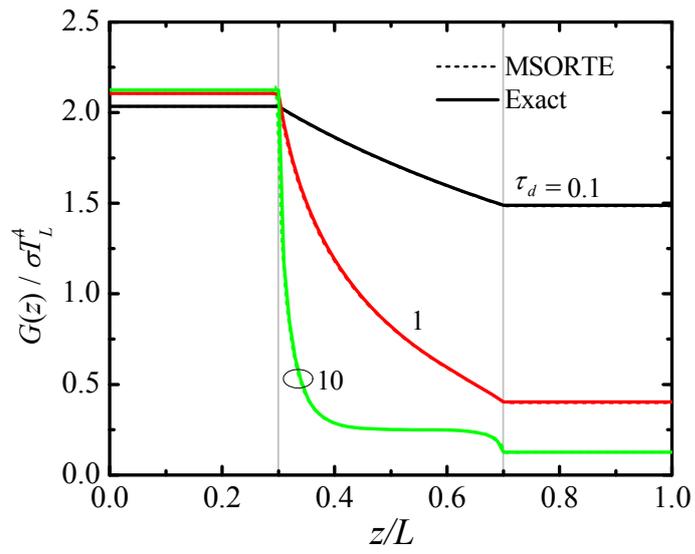

**Figure 10(b)**

**Authors: Zhao, Tan and Liu**

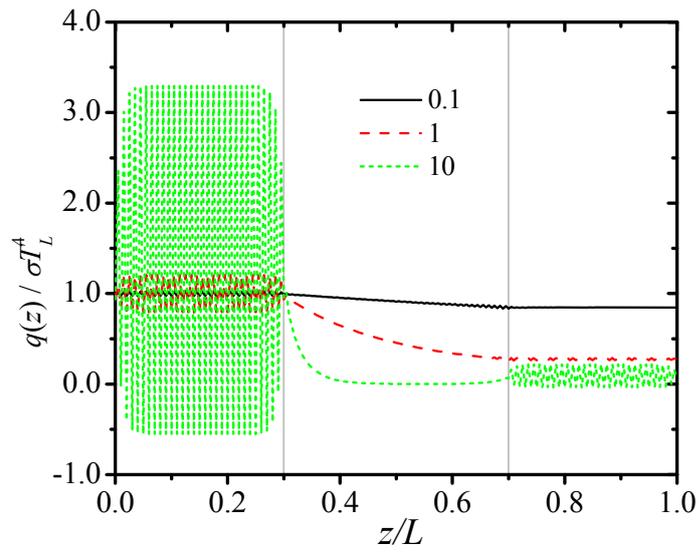

**Figure 11(a)**

**Authors: Zhao, Tan and Liu**

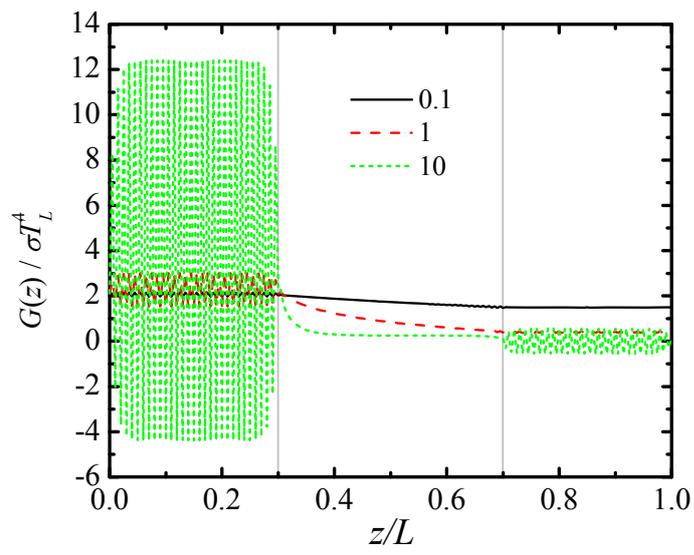

**Figure 11(b)**

**Authors: Zhao, Tan and Liu**

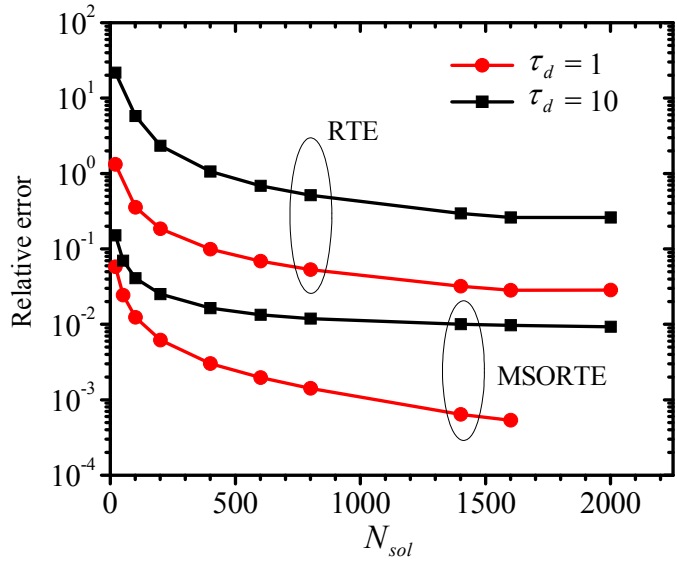

**Figure 12**

**Authors: Zhao, Tan and Liu**

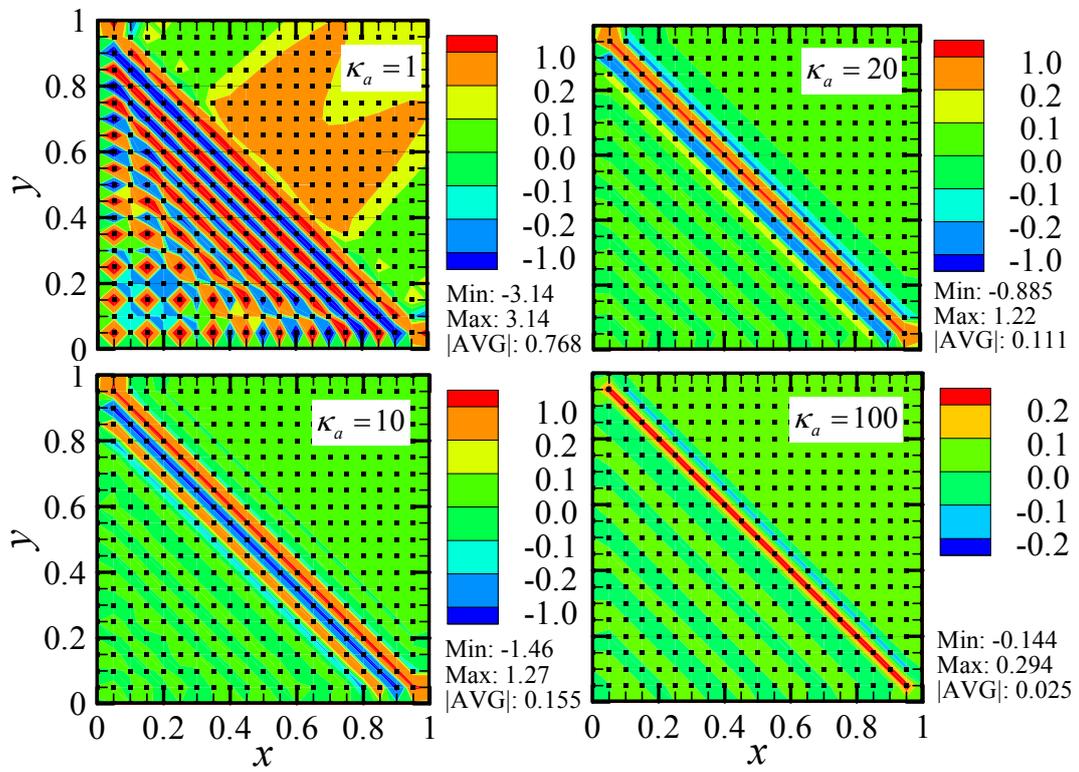

**Figure 13(a)**

**Authors: Zhao, Tan and Liu**

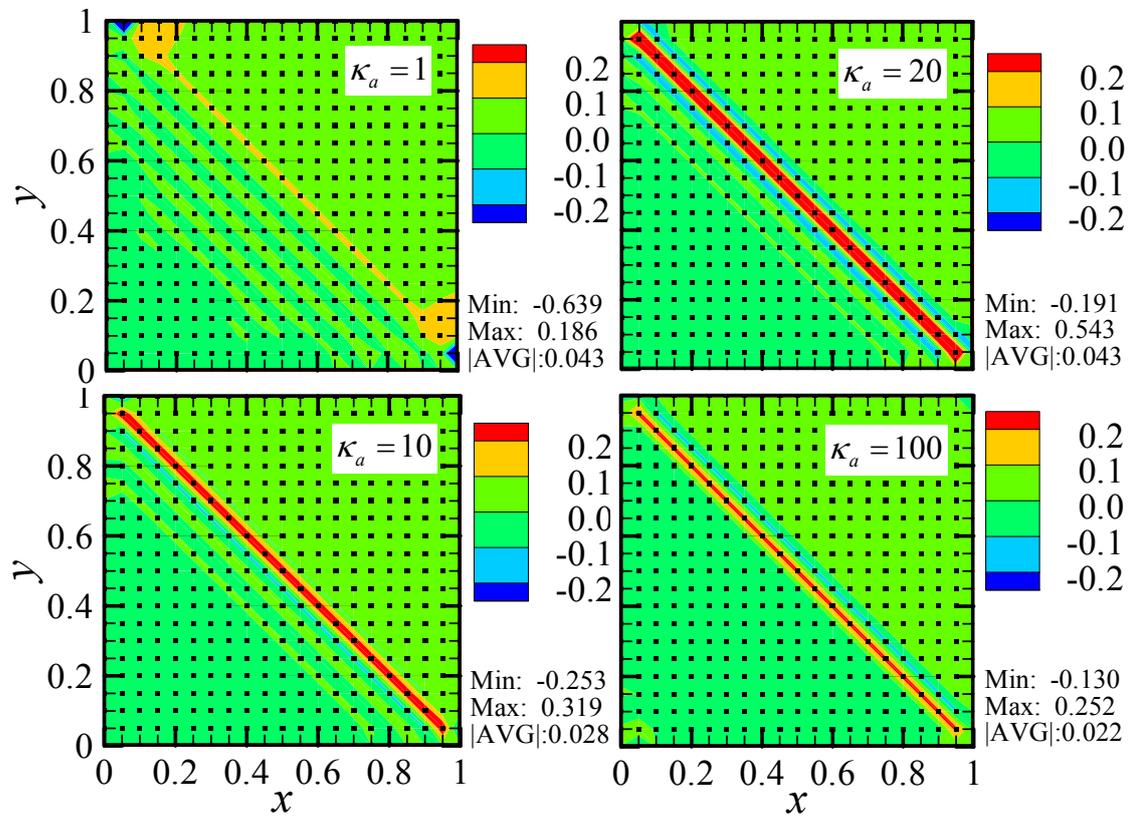

**Figure 13(b)**

**Authors: Zhao, Tan and Liu**

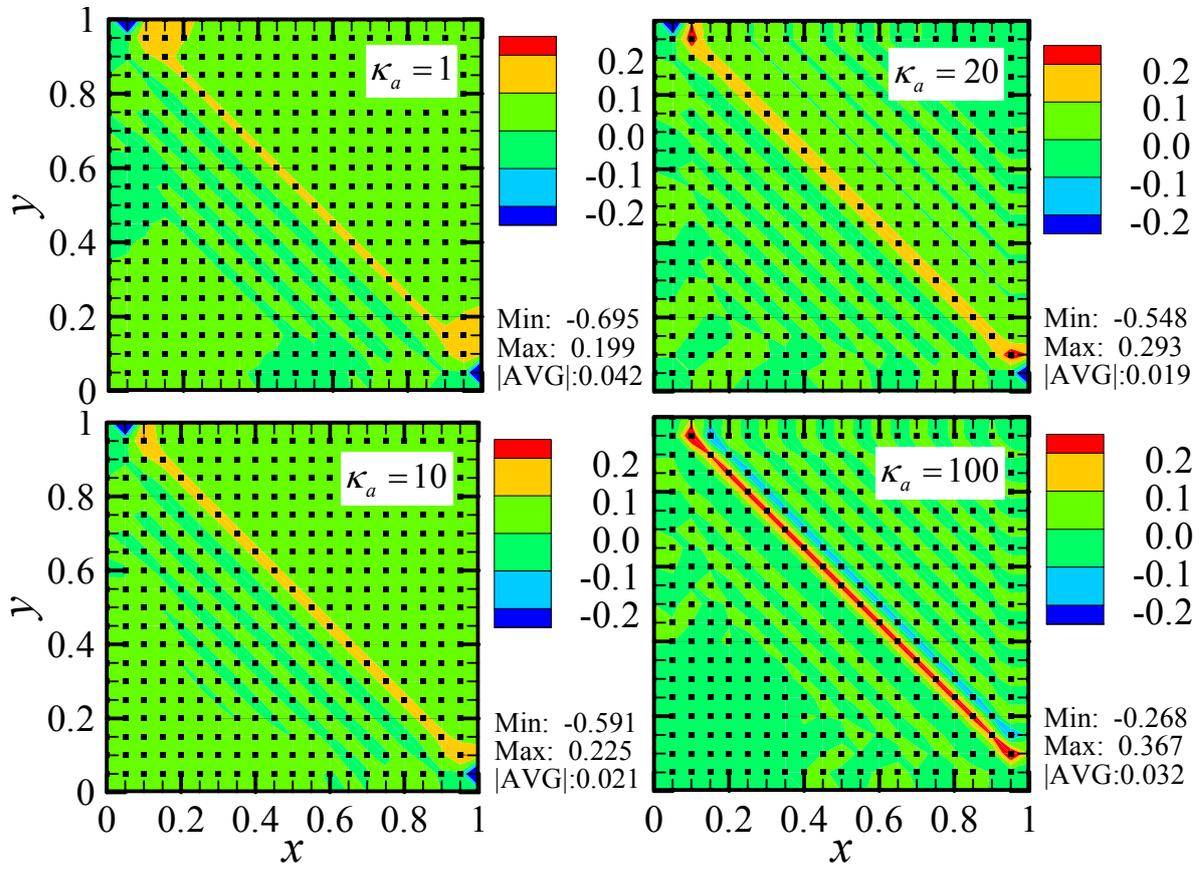

**Figure 13(c)**

Authors: Zhao, Tan and Liu

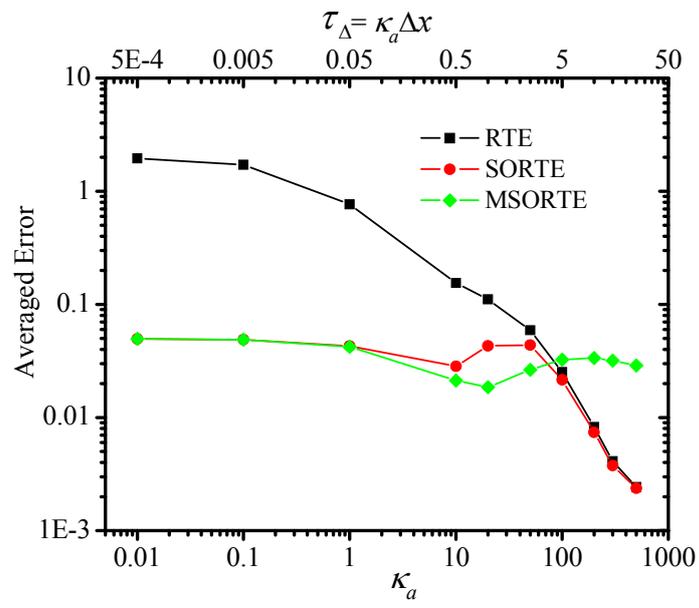

**Figure 14(a)**

**Authors: Zhao, Tan and Liu**

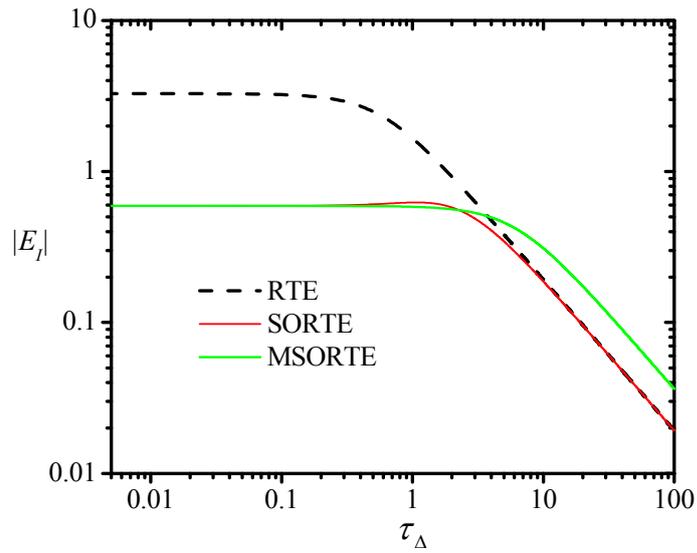

**Figure 14(b)**

**Authors: Zhao, Tan and Liu**

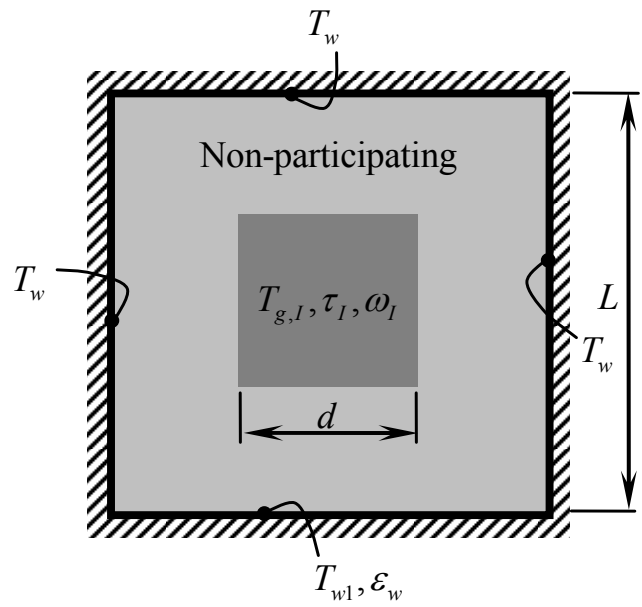

**Figure 15**

**Authors: Zhao, Tan and Liu**

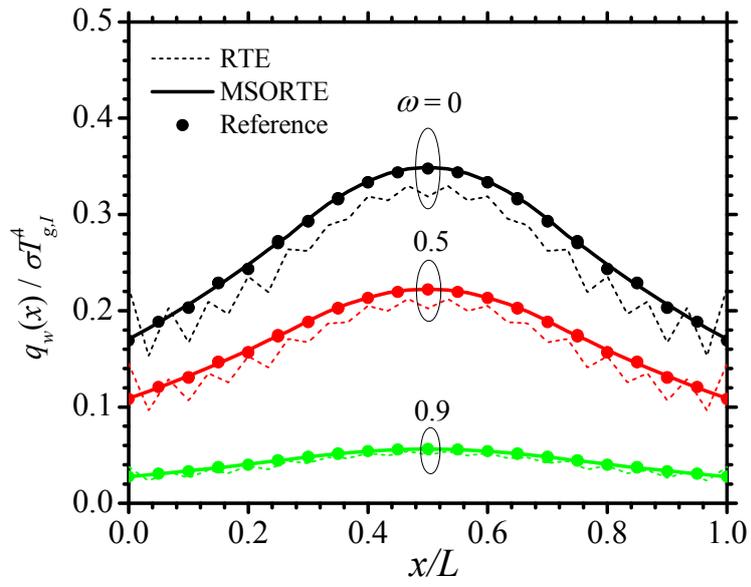

**Figure 16**

Authors: Zhao, Tan and Liu

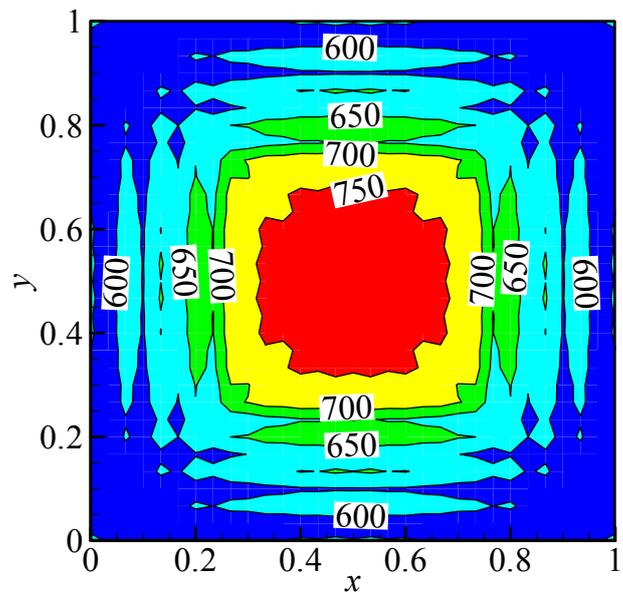

**Figure 17(a)**

**Authors: Zhao, Tan and Liu**

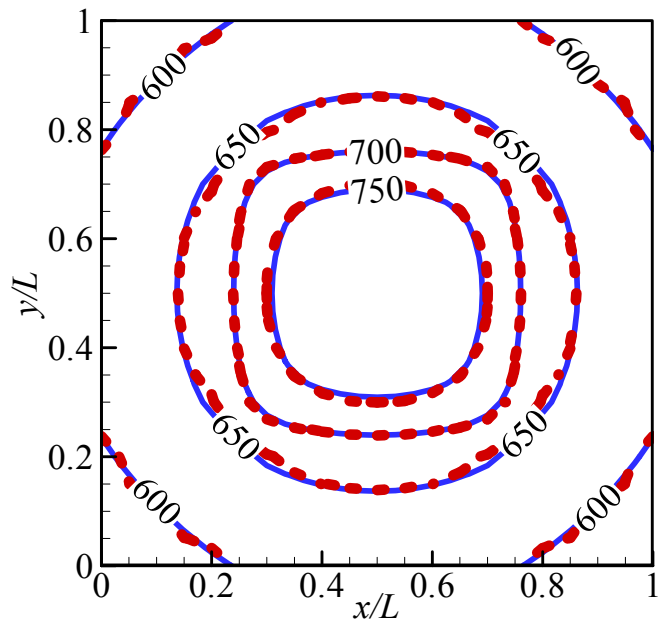

**Figure 17(b)**

Authors: Zhao, Tan and Liu

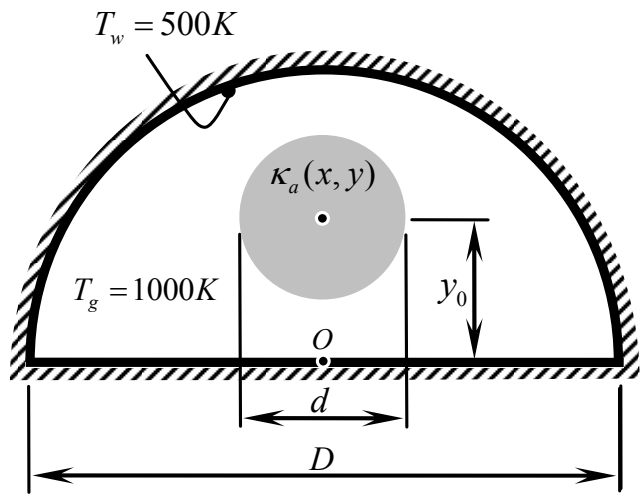

**Figure 18(a)**

**Authors: Zhao, Tan and Liu**

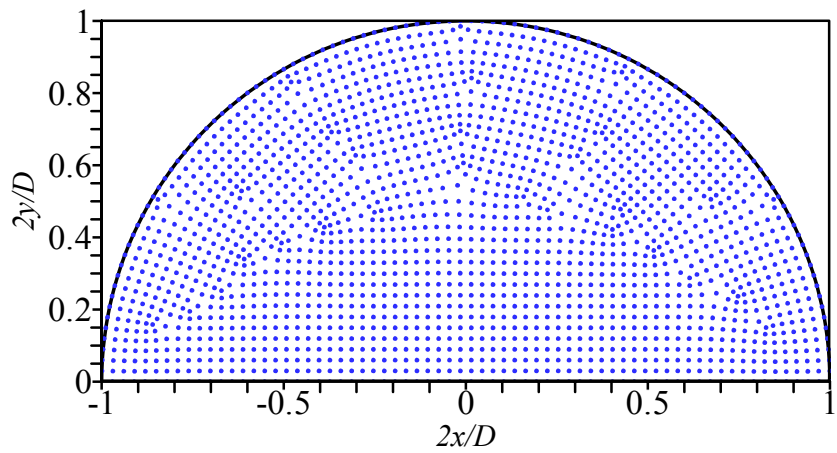

**Figure 18(b)**

**Authors: Zhao, Tan and Liu**

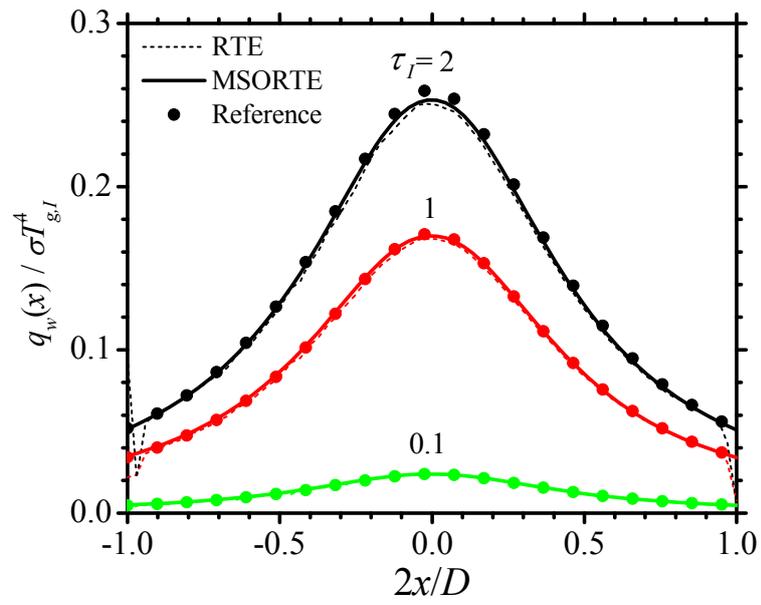

**Figure 19**

Authors: Zhao, Tan and Liu

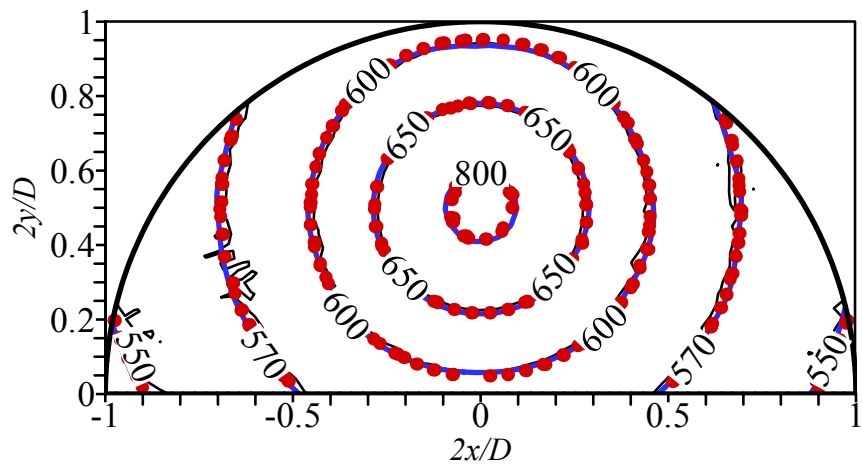

**Figure 20**

**Authors: Zhao, Tan and Liu**